\documentclass[article]{IEEEtran}
\usepackage{cite}

\usepackage{amsmath,amssymb,amsfonts}
\usepackage{algorithmic}
\usepackage{graphicx}
\usepackage{textcomp}
\usepackage{xcolor}
\usepackage{hyperref}
\usepackage{mathrsfs}
\usepackage{amsthm}
\usepackage{scalerel,stackengine}
\usepackage{paralist}
\usepackage[ruled,vlined,linesnumbered]{algorithm2e}

\usepackage{caption}
\usepackage{mathptm}
\usepackage{alltt}
\usepackage{epsfig}
\usepackage{verbatim}
\usepackage{fancybox}
\usepackage{subfigure}
\usepackage{url}
\usepackage{multirow}
\usepackage{setspace}
\usepackage{fancyhdr}
\usepackage{framed}
\usepackage{booktabs}
\usepackage{color}
\usepackage{wrapfig}

\usepackage{mathtools}
\mathtoolsset{showonlyrefs}
\usepackage{placeins}

\allowdisplaybreaks

\newtheorem{theorem}{Theorem}

\newtheorem{remark}[theorem]{Remark}
\newtheorem{example}[theorem]{Example}
\newtheorem{definition}[theorem]{Definition}

\newtheorem{assumption}[theorem]{Assumption}
\newtheorem{result}[theorem]{Result}

\begin{document}

	\title{Graph Signal Processing: Modulation, Convolution, and Sampling}
\author{John Shi, \IEEEmembership{Student Member, IEEE,} Jos\'e M.~F.~Moura, \IEEEmembership{Fellow, IEEE}%
\thanks{This material is based upon work partially funded and supported by the Department of Defense under Contract No. FA8702-15-D-0002 with Carnegie Mellon University for the operation of the Software Engineering Institute, a federally funded research and development center. This work is also partially supported by NSF grants CCF~1837607 and CCN~1513936.}%
\thanks{Department of Electrical and Computer Engineering, Carnegie Mellon University, Pittsburgh PA 15217 USA; [jshi3,moura]@andrew.cmu.edu.}}
	
	\maketitle
	
	\begin{abstract}
To analyze data supported by arbitrary graphs~$G$, DSP has been extended to Graph Signal Processing~(GSP) by redefining traditional DSP concepts like shift, filtering, and Fourier transform among others. This paper revisits modulation, convolution, and sampling of graph signals as appropriate natural extensions of the corresponding DSP concepts. To define these for both the vertex and the graph frequency domains, we associate with generic data graph~$G$ and its graph shift~$A$ a graph spectral shift~$M$ and a spectral graph $G_s$. This leads to a spectral GSP theory that parallels in the graph frequency domain the existing GSP theory in the vertex domain. The paper applies this to design and recovery sampling techniques for data supported by arbitrary directed graphs.
	\end{abstract}
	\vspace*{-.4cm}
	\section{Introduction}\label{sec:introduction}
	Traditional discrete signal processing (DSP) allows for the interpretation and analysis of well-ordered time and image signals. However, increasingly, data has an irregular structure that is better defined through a graph~$G$. Graph signal processing (GSP) extends traditional DSP to graph signals\textemdash data indexed by the nodes in~$G$. Applications include traffic data \cite{deriasilomar2015,deriFranzMoura2016dijkstra}, telecommunication networks, brain networks, or social media relationship networks, among many others, see these and further examples in \cite{jackson2010social,newman2010networks,easley2010networks}. For example, the features of a research paper in a citation network or the political leanings of hyperlinked blogs \cite{Adamic:2005} can be graph signals on a graph where each paper or each blog are represented by a node in the graph.

Two comments:
\begin{inparaenum}[(1)]
\item \label{inpara:GSP-DSP} We take GSP \cite{Sandryhaila:13,ShumanNFOV:13,Sandryhaila:14,Sandryhaila:14big,ortegafrossardkovacevicmouravandergheynst-2018} as an extension of DSP \cite{siebert-1986,oppenheimwillsky-1983,oppenheimschaffer-1989,mitra-1998} and Algebraic Signal Processing~(ASP) \cite{Pueschel:03a,Pueschel:05e,Pueschel:08a,Pueschel:08b,Pueschel:08c}, and as such GSP should as much as possible reduce to DSP and ASP when applied to the corresponding frameworks; and
    \item \label{inpara:lackspectralshift} in DSP it is well known how to shift signals in time and frequency, in ASP how to shift in space \cite{Pueschel:08b}, and in GSP how to shift a graph signal in the vertex domain \cite{Sandryhaila:13};  but neither ASP or GSP consider how to shift a signal in the spectral domain.
        \end{inparaenum}
Both points are relevant when we study modulation, convolution, and sampling\textemdash our main focus. From comment~\eqref{inpara:GSP-DSP}, we will look for graph modulation, graph convolution, and graph sampling as natural extensions to their DSP counterparts. On comment~\eqref{inpara:lackspectralshift}, we associate to the graph shift~$A$ a spectral graph shift~$M$ acting and defining in the graph spectral domain
\begin{inparaenum}[(1)]
\item a spectral graph $G_s$; and
    \item $M$-shift invariant spectral graph polynomial filters $P(M)$.
        \end{inparaenum}
          All three $M$, $G_s$, and $P(M)$ are relevant in their own right. For time signals, the spectral graph $G_s$ is equivalent (equal up to relabeling of the vertices) to the time graph~$G$ (same adjacency matrix~$A$), and this may be the reason why it has not appeared in the DSP literature. In GSP, in general, $M\neq A$, $G_s\neq G$, and $P(M)\neq P(A)$. The nodal and spectral graph shifts~$A$ and~$M$ play symmetric roles\textemdash for example,
           \begin{inparaenum}[(1)]
           \item modulation (pointwise product) in the nodal domain is filtering in the frequency domain with a polynomial filter $P(M)$; and
                \item filtering in the nodal domain with polynomial filter $P(A)$ is modulation with the graph filter frequency response in the graph spectral domain.
                     \end{inparaenum}

           The paper reviews GSP in section~\ref{sec:primerongsp}, defines spectral shift in section~\ref{sec:spectralshift}, considers graph impulses in section~\ref{sec:deltafunction}, studies convolution, filtering, and modulation in section~\ref{sec:modconvolution}. Section~\ref{sec:gspsampling} focus on GSP sampling, while section~\ref{sec:dspsampling} captures DSP sampling in the GSP sampling framework. Section~\ref{sec:connectionothersampling} relates ours to the work of others. Section~\ref{sec:conclusion} concludes the paper.          %
%

\textbf{Prior work}.\label{sec:previouswork}
 To study convolution and sampling, we consider a spectral shift~$M$ \cite{shimoura-asilomar2019}. Reference \cite{leus2017dual} has introduced previously a different graph shift, see section~\ref{sec:spectralshift} for connections between the two. ASP \cite{Pueschel:08a,Pueschel:08b} considers impulse signals in the algebraic context, but not the alternatives we discuss here. Convolution in the nodal domain has been studied as filtering (matrix-vector product) where the filter is defined by a polynomial filter \cite{Pueschel:08b,Pueschel:08b,Sandryhaila:13,Sandryhaila:14,ortegafrossardkovacevicmouravandergheynst-2018}, or by pointwise multiplication in the spectral domain \cite{ShumanNFOV:13}. In the paper,
  we study directly convolution of two graph signals (or filtering as convolution of an input graph signal and the graph impulse response that we define). This study shows the relevance of the spectral shift~$M$. The significance of considering in detail these concepts becomes apparent when we consider graph sampling. Graph sampling has been extensively studied, we will comment on a few of these papers and refer to them for a more complete review of the literature and  list of references. References \cite{pesenson2001sampling,pesenson2008sampling,pesenson2010sampling} consider the space of bandlimited graph signals (Paley-Wiener spaces) and establish that low-pass graph signals can be perfectly reconstructed from their values on some subsets of vertices (sampling sets). Sampling has received considerable attention in the GSP literature \cite{Narang:11,Narang:12,narang2013signal,anis2014towards,shomorony2014sampling,Jelena,gadde2015probabilistic,marques2015sampling,segarra2015interpolation,casey2015sampling,wang2015generalized,anis2016efficient,chen2016signal,sakiyama2016efficient,tsitsvero2016signals,segarra2016reconstruction,chamon2017greedy,xie2017design,anis2017critical,tremblay2017graph,Tanaka,jayawant2018distance,anis2018sampling,puy2018random,watanabe2018critically,sakiyama2019eigendecomposition,sakiyama2019two,guler2019robust}.
 These references address down- and up-spectral and vertex sampling, perfect, robust, greedy reconstruction, vertex domain eigenvector free sampling, interpolation of graph signals, sampling set selection, a probabilistic interpretation or a distance-based formulation of sampling, use graph sampling to solve sensor position selection, critical sampling for wavelet filterbanks, sampling of graph signals through successive local aggregations, uncertainty principles, among many other topics. Of particular relevance to our sampling work are \cite{Jelena,anis2016efficient,Tanaka}, and we will discuss relations to these in section~\ref{sec:connectionothersampling}. We emphasize that our main contribution is to provide a GSP sampling framework that parallels that of DSP sampling and showing the exact duality between vertex domain and spectral graph domain sampling, addressing the precise meaning of statements like ``vertex domain sampling cannot inherit the desired characteristics of the sampling in the graph frequency domain [.]'' \cite{sakiyama2019eigendecomposition} or ``[$\cdots$] in contrast to the classical case, the resulting signal in the graph frequency domain generally has a spectrum that cannot be separated into main and aliasing components even when the signal is bandlimited [.]'' \cite{sakiyama2019two}. We strive instead to emphasize the duality between vertex and graph spectral sampling.
 \vspace*{-.3cm}
\section{Primer in GSP}\label{sec:primerongsp}
\begin{figure}[hbt]
\vspace*{-.5cm}
		\includegraphics[height=3cm,width = 6cm]{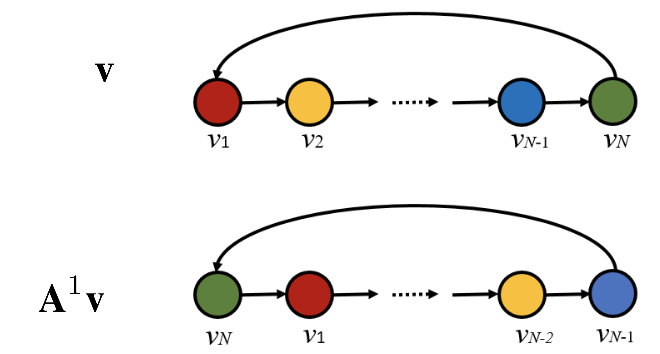}
		\caption{Ring graph: graph signal~$x$ and its shifted $A\cdot x$.}
		\label{fig1}
	\end{figure}

We cast DSP in the framework of GSP. A periodic time signal $\left\{x_0,x_1,\hdots,x_{N-1}\right\}$, period $N$, is defined on a ring graph of~$N$ nodes (top of Figure~\ref{fig1}). Each of the~$N$ signal samples, $x_0,x_1,\hdots,x_{N-1}$, is labelled by a node in the graph. Collect the signal in the $N \times 1$ vector, $x=[x_0,x_1,\hdots,x_{N-1}]^T$. The adjacency matrix for the ring graph is:
\begin{align}
\label{eqn:graphshiftA-1}
	A &{}= \begin{bmatrix}
	0  & 0 & 0 & \hdots & 0 & 1 \\
	1  & 0 & 0 &  \hdots & 0 & 0 \\
	\vdots  & \vdots & \ddots & \ddots & \vdots  & \vdots\\
	0  & 0 & 0 &  \ddots & 0 & 0 \\
	0  & 0 & 0 &  \hdots & 1 & 0 \\
	\end{bmatrix}
	\end{align}
	%
	 The adjacency matrix~$A$ in~\eqref{eqn:graphshiftA-1} is also the matrix representation of the DSP shift $z^{-1}$\textemdash the cyclic matrix. We shift the signal by multiplication by the shift~$A$ to get $A\cdot x$, shown at the bottom of Figure~\ref{fig1}.
	The Discrete Fourier Transform~(DFT) is found through the eigendecomposition of the shift matrix~$A$:
	\begin{align}\label{Adecomp}
	A &{}= \textrm{DFT}^\textrm{H} \cdot \Lambda \cdot \textrm{DFT}
	\end{align}
The diagonal matrix $\Lambda$ and the $\textrm{DFT}$ are
{\small
\begin{align}
\label{eqn:dspLambda}
\Lambda{}&=\textrm{diag}\left[\lambda_0\cdots\lambda_{N-1}\right], \: \: \: \lambda_k=e^{-j\frac{2\pi}{N}k},\:\:\: k=0\cdots (N-1)\\
\label{eqn:DFT1}
\textrm{DFT}{}&=\frac{1}{\sqrt{N}}\left[\begin{array}{llll}
1&e^{-j\frac{2\pi}{N}}&\cdots&e^{-j\frac{2\pi}{N}(N-1)}\\
\vdots&\vdots&&\vdots\\
1&e^{-j\frac{2\pi}{N}(N-1)}&\cdots&e^{-j\frac{2\pi}{N}(N-1)(N-1)}
\end{array}\right]\!\!.
\end{align}
}%
In~\eqref{eqn:DFT1} and below, all indices are \hspace*{-.2cm}$\mod\,N$. Since $\textrm{DFT}$ is unitary,  $\textrm{DFT}^H=\textrm{DFT}^{-1}$ ($H$ stands for transpose conjugate). The eigenvalues of~$A$ are $\lambda_k$, $k=0\cdots N-1$, and the eigenvectors $v_k$ (columns of $\textrm{DFT}^H$)
\begin{align}\label{eqn:keigenvectorA}
v_k{}&=\frac{1}{\sqrt{N}}\left[1 \,\,e^{j\frac{2\pi}{N}k}\,\cdots\,\, e^{j\frac{2\pi}{N}(N-1)}\right]^T, \,\,k=0\cdots N-1.
\end{align}
  \begin{remark}\label{rem:notation1}
       A subindex~$k$ may refer to the $k$th vector or the $k$th entry of a vector. The context should remove the ambiguity.
      \end{remark}
GSP extends DSP to indexing sets~$V$, the vertex set of arbitrary directed or undirected graph $G=(V,E)$, with~$E$ the set of edges of~$G$. The graph signal~$x\in\mathbb{C}^N$ assigns a data sample~$x_n$ to node~$n$, $n=0\cdots N-1$, of~$G$. Following \cite{Sandryhaila:13}, the graph shift is the adjacency matrix\footnote{Other authors consider other shifts, e.,g., the symmetric and positive definite graph Laplacian \cite{ShumanNFOV:13}, with real nonnegative eigenvalues but restricted to undirected graphs, or unitary variations of~$A$ that sacrifice locality \cite{giraultgoncalvesfleury-2015,gavilizhang-2017}.} $A$. We shift~$x$ by applying shift~$A$, i.e., $A\cdot x$. The graph Fourier Transform~(GFT) is defined through the eigendecomposition of the shift~$A$
	\begin{align}\label{Adecompgsp}
	A {}&= \textrm{GFT}^\textrm{-1} \cdot \Lambda \cdot \textrm{GFT}
	\end{align}
%
\begin{assumption}\label{assp:diagonalizableA} Shift~$A$ has distinct eigenvalues.
\end{assumption}
Under assumption~\ref{assp:diagonalizableA}, generic asymmetric~$A$ is diagonalizable. For the more general case of repeated eigenvalues and/ or non diagonalizable shifts, see \cite{Sandryhaila:13} and \cite{derimoura-2017}. The eigenvalues are the graph frequencies\footnote{In DSP, frequencies are commonly $\Omega_k=\frac{2\pi}{N}k=-\frac{1}{2\pi j}\ln e^{-j\frac{2\pi}{N}k}$, not $\lambda_k$.}  and the columns of $\textrm{GFT}^\textrm{-1}$ are the eigenvectors of~$A$ referred to as spectral components, playing the role of the harmonics in time signals. In GSP, filtering in the vertex domain is defined as a matrix vector multiplication. 	
 \begin{align}\label{convtime}
	P(A) \cdot x
	\end{align}
	where $x$ is the graph signal, In~\eqref{convtime}, the filter is shift invariant and so $P(A)$ is a polynomial of the shift, $A$ \cite{Sandryhaila:13}.
	
	While \cite{Sandryhaila:13,Sandryhaila:14,Sandryhaila:14big} provide fundamental GSP operations, they do not define shift in the frequency domain, convolution in the frequency domain, or modulation. Furthermore, while they define convolution in the vertex domain through filtering (matrix (filter) vector (graph signal) product), they do not define convolution of two graph signals, nor find the graph filter $P(A)$ given its graph impulse response. To address these issues, we consider shifting in the graph spectral domain first.
\vspace*{-.3cm}
	\section{Shift in Graph Spectral Domain}
	\label{sec:spectralshift}
	We start by considering the effect of shifting a signal in the frequency domain from which we propose a spectral shift~$M$ acting on the graph spectral domain \cite{shimoura-asilomar2019}. In \cite{leus2017dual}, the authors define a different spectral shift~$M^\prime$, see below, that they require to satisfy a number of properties; for example, permutation invariance that is unfortunately seldom verified.

We derive the GSP graph spectral shift~$M$ from first principles. It will play a significant role in modulation and filtering in the spectral domain, see section~\ref{sec:modconvolution}. To determine the spectral shift, we first consider the ring graph and DSP.
	\vspace*{-.6cm}
	\subsection{DSP: Spectral shift~$M$}\label{subsec:dspspectralshift}
	In DSP, the following property holds:
	\begin{align}\label{shiftprop}
	e^{j\frac{2\pi}{N}m_0k}x_k \xrightarrow{\mathcal{F}} \widehat{x}_{m-m_0},
	\end{align}
	where $\widehat{x}_m$ is the $m$th Fourier coefficient of the time signal. Then, \eqref{shiftprop} shows that shifting in the frequency domain $\widehat{x}_m$ by $m_0$ multiplies the signal sample $x_k$ by $\left(\lambda_k^*\right)^m$, the complex conjugate of the eigenvalue, raised to the power $m_0$. Letting $\Lambda^*  = \textrm{diag}\left(e^{j\frac{2\pi}{N}k}\right)$ and stacking the time samples in vector~$x$ and the shifted graph Fourier components in a vector, we get
	\begin{align}\label{shiftprop-2}
	\left(\Lambda^*\right)^{m_0} x \xrightarrow{\mathcal{F}}
\left[\begin{array}{ccc}
\widehat{x}_{-m_0}&
\cdots&
\widehat{x}_{N-1-m_0}
\end{array}\right]^T.
	\end{align}
We now define the spectral shift~$M$. Consider the eigendecomposition of~$A$ in~\eqref{Adecomp}, with the $\textrm{DFT}$ as defined in~\eqref{eqn:DFT1}.
\begin{definition}[DSP: Spectral shift~$M$]\label{def:dspspectralshiftM}
The spectral shift~$M$ is
	\begin{align} \label{Meqn}
	M {}&= \textrm{DFT} \cdot \Lambda^* \cdot \textrm{DFT}^\textrm{H}
	\end{align}
\end{definition}
	We show~$M$ acts as spectral shift.
%
%
 For~$x$ and $\widehat{x} = \textrm{DFT}\cdot x$
	\begin{align} \label{shifteqn}
	M\widehat{x}= \textrm{DFT} \cdot \Lambda^* \cdot \textrm{DFT}^\textrm{H}& \cdot \textrm{DFT}\cdot x = \textrm{DFT}\cdot\Lambda^* \cdot x\\
\label{shifteqn-2}
\Lambda^* x&\xrightarrow{\mathcal{F}} M\cdot\widehat{x}
	\end{align}
\begin{result}[DSP: $A=M$] \label{res:shiftsAM} \label{AMequal}
For the time shift~\eqref{eqn:graphshiftA-1} and direct cyclic graph with diagonalization~\eqref{Adecomp} with the $\textrm{DFT}$~\eqref{eqn:DFT1}
\begin{align}\label{eqn:shiftsAM}
A{}&=M.
\end{align}
\end{result}
To prove the result, conjugate~$M$ in~\eqref{Meqn} to get $M^*=A$ and then realize that in DSP~$A$ is real valued and so $M=A$.
	%
%
%
\vspace*{-.5cm}
	\subsection{GSP: Spectral shift~$M$ and spectral graph}\label{subsec:gspspectralshift}
	We adjust definition~\ref{def:dspspectralshiftM} to define the GSP spectral shift \cite{shimoura-asilomar2019}.
\begin{definition}[GSP: Spectral shift~$M$]\label{def:gspspectralshiftM}
Given the diagonalization~\eqref{Adecompgsp} of the graph shift~$A$, the spectral shift~$M$ is
	\begin{align} \label{eqn:Mgeqn}
	M {}&= \textrm{GFT} \cdot \Lambda^* \cdot \textrm{GFT}^{-1}
	\end{align}
where~$\textrm{GFT}$ and $\Lambda^*$ are the~$\textrm{GFT}$ and the conjugate of the diagonal matrix of eigenvalues of the shift~$A$ given in~\eqref{Adecompgsp}.
\end{definition}
While for DSP, $M=A$, in GSP~$M$ may not equal~$A$. They play twin roles. Besides being shifts, they define, as adjacency matrices, graphs: shift~$A$ defines the (data) graph~$G$ whose node~$n$ indexes the data sample $x_n$, while the spectral shift~$M$ defines a new graph, the \textit{spectral} graph $G_s$, whose node~$m$ indexes the graph Fourier coefficient $\widehat{x}_m$ of the data. We emphasize that each node~$m$ of the spectral graph stands for a graph frequency, say, $\lambda_m$, $m=0\cdots N-1$.

\begin{remark}\label{rem:twodefspshift}
Reference~\cite{leus2017dual} defines the spectral shift with $\Lambda$ rather than $\Lambda^*$ in~\eqref{eqn:Mgeqn}. This is undesirable. For example, for the time shift~\eqref{eqn:graphshiftA-1} (directed cycle graph) with diagonalization~\eqref{Adecomp} and the $\textrm{DFT}$ given in~\eqref{eqn:DFT1}, by result~\eqref{res:shiftsAM} our definition~\ref{def:dspspectralshiftM} leads to the equality
$A=M$ as desired. In contrast, with the definition in~\cite{leus2017dual}, the time and spectral shifts~$M^\prime=A^T\neq A$, reversing the direction of the cycle graph.
\end{remark}

\textit{$M$ is structural unique.} But the analogies stop short. It is well known that the spectral modes, eigenvectors of~$A$, are not unique. For the implications in GSP of this nonunicity see \cite{derimoura-2017}. Under assumption~\ref{assp:diagonalizableA} on~$A$, the spectral modes (eigenvectors) of~$A$ form a complete basis, and so they are unique up-to-scaling\textemdash multiplication of the matrix of eigenvectors on the right by a diagonal matrix~$C=\textrm{diag}\left[c_0\cdots c_{N-1}\right]^T$, $\forall c_n\neq0$, with possibly different diagonal entries. This of course leaves the diagonalizable matrix~$A$ and its graph~$G$ invariant
\begin{align}
\label{eqn:Ainvscaling}
A=\textrm{GFT}^H \cdot C\cdot\Lambda \cdot C^{-1}\cdot\textrm{GFT},
\end{align}
since two diagonal matrices commute. From~\eqref{eqn:Ainvscaling}, $C^{-1}\cdot\textrm{GFT}$ is still a graph Fourier transform, but different for different $C\neq 0$. The corresponding~$M$, labeled $M_C$, is
\begin{align}
\label{eqn:Mscaling}
M_C=C^{-1} \cdot \textrm{GFT}\cdot\Lambda^* \cdot \textrm{GFT}\cdot C,
\end{align}
which is different from~$M$ in~\eqref{eqn:Mgeqn}. The relation between the spectral shifts and spectral graphs $G_s$ for the same~$A$ is next.
\begin{result}[Structural invariance of the spectral graph~$G_s$]\label{res:unicityspgraph}
Given a shift~$A$, any two corresponding spectral shifts $M_1$ and $M_2$ are conjugate by an invertible diagonal matrix~$C$
\begin{align}\label{eqn:Mconjugate}
M_2{}&=C^{-1} \cdot M_1\cdot C
\end{align}
 with their spectral graphs~$G_{s_1}$ and~$G_{s_2}$ structurally equivalent.
\end{result}
In result~\ref{res:unicityspgraph}, the structure of a graph is defined by the nonzero entries in its spectral shift~$M$ (set of edges of the graph). To prove result~\ref{res:unicityspgraph}, observe that equation~\eqref{eqn:Mconjugate} follows from~\eqref{eqn:Mscaling} and $C\neq 0$. The structural equivalence follows because conjugation of a spectral shift~$M_1$ by an invertible diagonal matrix~$C$ rescales (differently) the entries of~$M_1$, except the diagonal entries of~$M_1$ and its zero entries. So, result~\ref{res:unicityspgraph} leaves the structure of~$M_1$ and of~$G_{s_1}$ invariant.  But the conjugated spectral graphs~$G_{s_1}$ and~$G_{s_2}$ have different weights.
\begin{example}[Star graph]\label{exp:stargraphM}
	\label{subsec:propertiesM}
Consider a star graph with adjacency
\begin{equation} \label{eqn:starA}
    A =
\begin{bmatrix}
 0 & 1 & 1 & \hdots & 1\\
 \vspace*{-.2cm} 1 & 0 & 0 & \hdots & 0\\
\vdots & \vdots & \vdots & \ddots & \vdots \\
 1 & 0 & 0 & \hdots & 0\\
\end{bmatrix}
\end{equation}
%
The eigenvalues of $A$ are $\pm\sqrt{N-1}$ with multiplicity~1 and~0 with algebraic and geometric multiplicity~$N-2$.
%
We get
\begin{align} \label{eqn:stargfti}
    \textrm{GFT}^{-1} {}&=
    {\scriptsize
 \phantom{\frac{1}{N-1}}\:\:\:\:\:
\left[
\begin{array}{cccccc}
 \sqrt{N-1} & \hspace{-.2cm}-\sqrt{N-1} & 0 & 0 & \hdots & 0\\
 1 & 1 & \hspace{-.2cm}-1 & \hspace{-.2cm}-1 &  \hdots & \hspace{-.2cm}-1\\
 1 & 1 & 1 & 0& \hdots & 0\\
  0 & \hspace{-.2cm}-1 & \hspace{-.2cm}-1 & N-2 &  \cdots & \hspace{-.2cm}-1\\
 \vspace*{-.2cm} 1 & 1 & 0 & 1 &\hdots & 0\\
 \vdots & \vdots & \vdots & \vdots & \ddots & \vdots \\
 1 & 1 & 0 & 0 &  \hdots & 1\\
\end{array}
\right]
}
\\
\label{eqn:stargft}
\textrm{GFT}{}& =
{\scriptsize
\frac{1}{N-1}
\left[
\begin{array}{cccccc}
 \hspace{-.2cm}\phantom{-}\frac{\sqrt{N-1}}{2} & \phantom{-}\frac{1}{2} & \phantom{-}\frac{1}{2}  & \phantom{-}\frac{1}{2}  & \cdots & \phantom{-}\frac{1}{2} \\
 \hspace{-.2cm}-\frac{\sqrt{N-1}}{2} & \phantom{-}\frac{1}{2} & \phantom{-}\frac{1}{2}  & \phantom{-}\frac{1}{2}  & \cdots & \phantom{-}\frac{1}{2} \\
  0 & -1 & -1 & N-2 &  \cdots & -1\\
 \vspace*{-.2cm}0 & -1 & N-2 & -1 &  \cdots & -1\\
 \vdots & \vdots & \vdots & \vdots & \ddots & \vdots \\
 0 & -1 & -1 & -1 &  \cdots & N-2\\
\end{array}
\right]
}
\\
\label{eqnstargft2}
  A{}  &= \text{GFT}^{-1} \textrm{diag}{\left[\sqrt{N-1}, -\sqrt{N-1}, 0, 0, \hdots, 0\right]} \text{GFT}\\
%
    M &= 
 \textrm{GFT}\phantom{^{-1}}
  \textrm{diag}
  {\scriptsize
  \left[\sqrt{N-1}, -\sqrt{N-1}, 0, 0, \hdots, 0\right]
  } \textrm{GFT}^{-1}
  \nonumber \\
&= \frac{1}{\sqrt{N-1}}
{\scriptsize
\left[
\begin{array}{cccccc}
 \frac{N-2}{2} & \hspace{-.2cm}-\frac{N}{2} & \frac{1}{2}  & \frac{1}{2}  & \hdots & \frac{1}{2} \\
\hspace{-.2cm} -\frac{N}{2} & \frac{N-2}{2} & \frac{1}{2}  & \frac{1}{2}  & \hdots & \frac{1}{2} \\
1 & 1 &\hspace{-.2cm} -1 & \hspace{-.2cm}-1& \hdots & \hspace{-.2cm}-1\\
 0 & \hspace{-.2cm}-1 & \hspace{-.2cm}-1 & N-2 &  \cdots &\hspace{-.2cm} -1\\
 \vspace*{-.2cm}1 & 1 & \hspace{-.2cm}-1 & \hspace{-.2cm}-1& \hdots & \hspace{-.2cm}-1\\
 \vdots & \vdots & \vdots & \vdots & \ddots & \vdots \\
1 & 1 & \hspace{-.2cm}-1 & \hspace{-.2cm}-1& \hdots & \hspace{-.2cm}-1\\
\end{array}
\right]
} \label{eqn:starM}
\end{align}

Using Equation \eqref{eqn:starA} and \eqref{eqn:starM} with $N = 5$ yields
\begin{align} 
A &=
{\scriptsize
\left[
\begin{array}{ccccc}
 0 & 1 & 1 & 1 & 1\\
 1 & 0 & 0 & 0 & 0\\
 1 & 0 & 0 & 0 & 0\\
 1 & 0 & 0 & 0 & 0\\
 1 & 0 & 0 & 0 & 0\\
\end{array}
\right]
},\:
M = \frac{1}{2}
{\scriptsize
\left[
\begin{array}{ccccc}
 \frac{3}{2} & \hspace{-.2cm}-\frac{5}{2} & \frac{1}{2} & \frac{1}{2} & \frac{1}{2}\\
\hspace{-.2cm}-\frac{5}{2} &  \frac{3}{2} &  \frac{1}{2} & \frac{1}{2} & \frac{1}{2}\\
 1 & 1 & \hspace{-.2cm}-1 & \hspace{-.2cm}-1 & \hspace{-.2cm}-1\\
 1 & 1 & \hspace{-.2cm}-1 & \hspace{-.2cm}-1 & \hspace{-.2cm}-1\\
 1 & 1 & \hspace{-.2cm}-1 & \hspace{-.2cm}-1 & \hspace{-.2cm}-1\\
\end{array}
\right]
}
\nonumber
\end{align}
The graphs corresponding to~$A$ and~$M$ are shown in Figure~\ref{fig:star}.
\vspace*{-.6cm}
\begin{figure}[hbt]
    \centering
    \includegraphics[width=6cm,keepaspectratio]{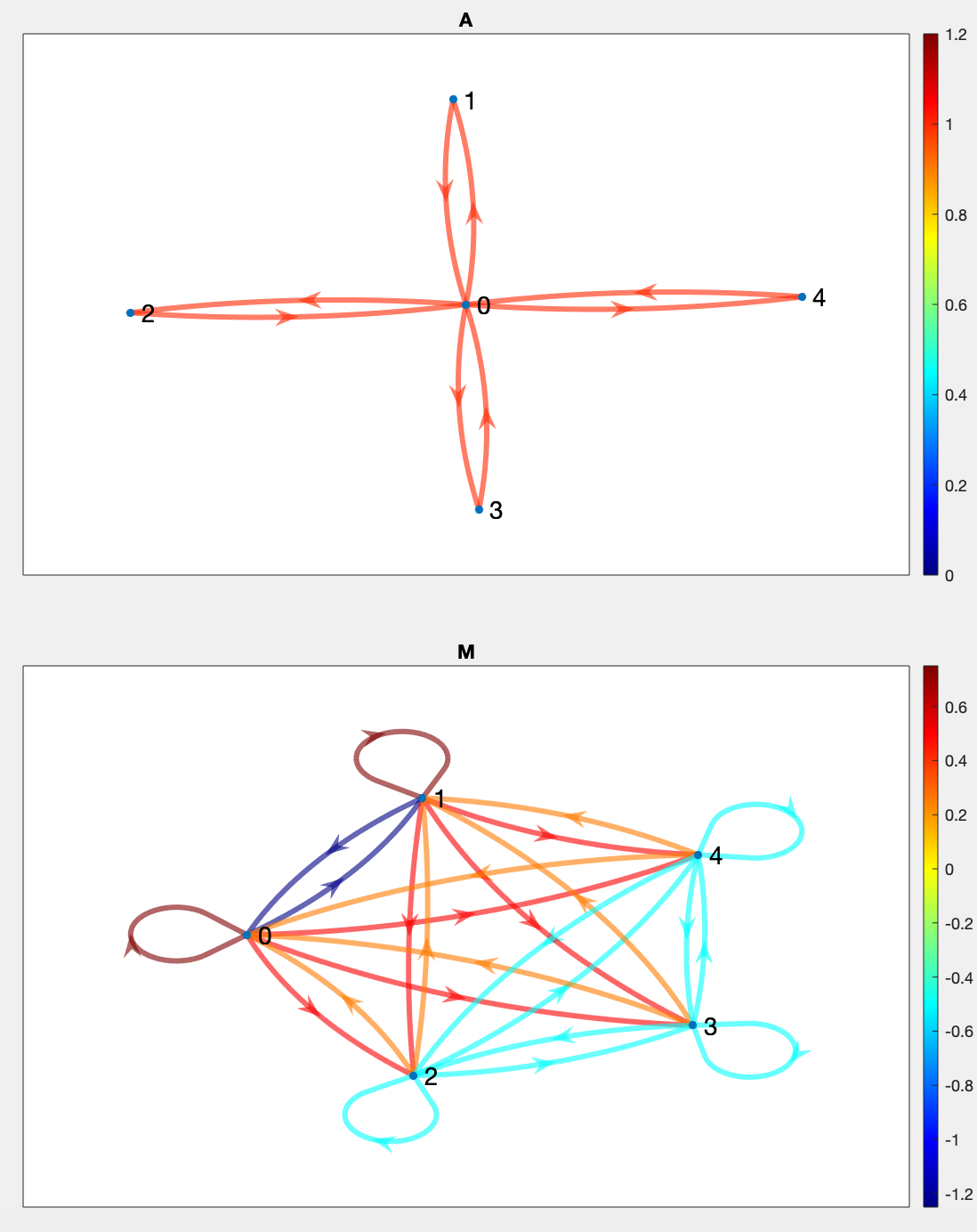}
    \caption{Star graph: Shifts~$A$ and~$M$.}
    \label{fig:star}
\end{figure}
\end{example}
\vspace*{-1.1cm}
	\section{Delta Functions in GSP}\label{sec:deltafunction}
	We consider the delta function and its shifts in GSP.
	Let $e_n$, $n=0\cdots N-1$, be a zero vector except entry~$n$ equals one\footnote{Recall that we number entries in this paper from~0, so $e_0=[1\,0\cdots0]^T$.}. In DSP, the delta function and its DFT are $\delta_0 = [1\,0 \hdots 0]^T=e_0$ and  $\widehat{\delta}_0 = \frac{1}{\sqrt{N}}[1\,1 \hdots 1]^T$\textemdash impulsive in time and flat in frequency. The DSP $n$th-time shifted delta is
\begin{align}\label{eqn:dsptimeshifteddelta}
\delta_n\!\!{}&=\!A^n\cdot\delta_0\!=\!e_n\!\xrightarrow{\mathcal{F}}\!\widehat{\delta}_n\!\!= \!\!
\frac{1}{\sqrt{N}}\lambda^n\!\!=\!\!
\frac{1}{\sqrt{N}}
\!\!
\left[\!\!\!\!\begin{array}{c}
\lambda_0^n\\
\lambda_1^n\\
\vdots\\
\lambda_{N-1}^n
\end{array}\!\!\!\!\right]
\!\!\!\!=\!\!\!\!
\frac{1}{\sqrt{N}}\!\!\!\!
\left[\!\!\!\!\!\!\begin{array}{c}
1\\
e^{-j\frac{2\pi}{N}n}\\
\vdots\\
e^{-j\frac{2\pi}{N}(N-1)n}
\end{array}\!\!\!\!\!\!\right]
\end{align}
where $\lambda^n$ is a vector whose entries are the graph frequencies raised to power~$n$. The DSP $n$th-spectral shifted delta is
\begin{align}\label{eqn:dspspectralshifteddelta}
\widehat{\delta}_n{}&=M^n\widehat{\delta}_0= \textrm{GFT}\,\Lambda^{*^n}\textrm{GFT}^{-1}\widehat{\delta}_0 =\textrm{GFT}\,\Lambda^{*^n}\delta_0 =\textrm{GFT}\,\delta_0=\widehat{\delta}_0.
\end{align}
The fourth equality follows because the first diagonal entry of $\Lambda^*$ is~$1$. It is intuitively pleasing that $\widehat{\delta}_n=\widehat{\delta}_0$ since circular shifts of a flat $\widehat{\delta}_0$ obtain the same function.

	In GSP, the delta function is either impulsive in the vertex domain or flat in the graph spectral domain, but not both in general. We consider briefly both choices.

\textit{Graph delta impulsive in the vertex domain $\delta^{(v)}_0=e_0$}: Get
\begin{align}
\label{eqn:GSPimpulseshiftdeltav}
\widehat{\delta^{(v)}_0}=\textrm{GFT}\cdot e_0=y_0{}&\Longrightarrow \delta^{(v)}_n=A^n\cdot\delta^{(v)}_0=a^{(n)}_0\\
\label{eqn:GSPimpulseshiftdeltav-3}
\widehat{\delta^{(v)}_n}=\widehat{A^n\cdot\delta^{(v)}_0}{}&=\Lambda^n\cdot y_0 =\mathcal{Y}_0\cdot\lambda^n\\
\label{eqn:GSPimpulseshiftdeltav-6}
\Longrightarrow \delta^{(v)}_n=\textrm{GFT}^{-1}{}&\left(\Lambda^n\cdot y_0\right)=\textrm{GFT}^{-1}\left(\mathcal{Y}_0\cdot \lambda^n\right)
\\
\label{eqn:GSPimpulseshiftdeltav-2}
\textrm{GFT}^{-1}{}&\left(M^n\cdot\widehat{\delta^{(v)}_0}\right)= \Lambda^{*^n}\cdot e_0=\lambda_0^{*^n}\cdot e_0
\end{align}
where $a^{(n)}_0$ is the $0$th-column of $A^n$, $y_0$ is the first column of the $\textrm{GFT}$, $\mathcal{Y}_0=\textrm{diag}\left[y_0\right]$, and $\lambda^n=\left[\lambda_0^n\cdots\lambda_{N-1}^n\right]^T$.  Note $\Lambda^n\cdot y_0 =\mathcal{Y}_0\cdot\lambda^n=\Lambda^n\mathcal{Y}_0\cdot 1$ because both $\Lambda^n$ and $\mathcal{Y}_0$ being diagonal commute and a diagonal times~$1$ gives the vector of the diagonal entries. Interestingly, it is the spectral shifted delta $M^n\,\widehat{\delta^{(v)}_0}$ see~\eqref{eqn:GSPimpulseshiftdeltav-2} that is impulsive delta in the vertex domain, not the vertex shifted delta $A^n\delta^{(v)}_0$ see~\eqref{eqn:GSPimpulseshiftdeltav-6}. Define the impulse matrix~$D^{(v)}$ and its GFT $\widehat{D^{(v)}}$; using~\eqref{eqn:GSPimpulseshiftdeltav} and~\eqref{eqn:GSPimpulseshiftdeltav-3}, get  
\begin{align}
\label{eqn:matriximpv}
D^{(v)}{}&\triangleq
\left[\begin{array}{cccc}
\delta_0^{(v)}&\delta_1^{(v)}&\cdots&\delta_{N-1}^{(v)}
\end{array}\right]
\\
\label{eqn:matriximpv-6}
{}&=\left[\begin{array}{cccc}
a_0^{(0)}&a_0^{(1)}&\cdots&a_{0}^{(N-1)}
\end{array}\right]\\
\label{eqn:matriximpv-2}
\widehat{D^{(v)}}{}&=\textrm{GFT} \cdot D\\
\label{eqn:matriximpv-3}
{}&= \frac{1}{\sqrt{N}}\phantom{\mathcal{Y}_0}\left[\begin{array}{cccc}
\widehat{\delta_0^{(v)}}&\widehat{\delta_1^{(v)}}&\cdots&\widehat{\delta_{N-1}^{(v)}}
\end{array}\right]\\
\label{eqn:matriximpv-4}
{}&= \frac{1}{\sqrt{N}}\mathcal{Y}_0 \left[\begin{array}{cccc}
\lambda^0&\lambda^1&\cdots&\lambda^{N-1}
\end{array}\right]\\
\label{eqn:matriximpv-5}
{}&= \underbrace{\mathcal{Y}_0\frac{1}{\sqrt{N}} \left[\begin{array}{cccc}
1&\lambda_0&\cdots&\lambda_0^{N-1}\\
\vdots&&&\vdots\\
1&\lambda_0^{N-1}&\cdots&\lambda_{N-1}^{N-1}
\end{array}\right]}_{\mathcal{V}_{\lambda}}
\end{align}
where $\mathcal{V}_{\lambda}$  is the (normalized) Vandermonde matrix \cite{gantmacher1959matrix} of the eigenvalues of~$A$ multiplied by $\mathcal{Y}_0$.
\begin{assumption}[Nonzero entries of first column $y_0$ of $\textrm{GFT}$]
\label{assp:nonzeroy0}
All the entries of $y_0$ are nonzero.
\end{assumption}
If Assumption~\ref{assp:nonzeroy0} holds, the diagonal matrix $\mathcal{Y}_0$ is invertible. For example, vector~$1$ is an eigenvector of row-stochastic matrices (nonnegative entries). These matrices are the shift for a very broad class of directed graphs (digraphs).  Further, if Assumption~\ref{assp:diagonalizableA} also holds, $A$ has distinct eigenvalues, and so $\mathcal{V}_{\lambda}$ is invertible. Then under these two assumptions, $D$ and $\widehat{D}$ are full rank and invertible. A quick check of~\eqref{eqn:matriximpv} through~\eqref{eqn:matriximpv-5}, for the DSP and the DFT, $y_0=\frac{1}{\sqrt{N}} 1$, $\mathcal{Y}_0=\frac{1}{\sqrt{N}} I$, where~$I$ is the $N$-dimensional identity, and  with the $\textrm{DFT}$ given in~\eqref{eqn:DFT1}
\begin{align}
\label{eqn:matriximpvDFT}
\widehat{D^{(v)}}=\mathcal{V}_{\lambda,{\scriptsize\textrm{DFT}}}=\textrm{DFT}.
\end{align}

\textit{Graph delta flat in the spectral domain $\widehat{\delta^{(s)}_0}=\frac{1}{\sqrt{N}}1$}: Get
\begin{align}
\label{eqn:GSPimpulseshiftdeltas-3}
\delta^{(s)}_0{}&=\frac{1}{\sqrt{N}}\sum_{k=0}^{N-1}v_k
\\
\label{eqn:GSPimpulseshiftdeltas-4}
\widehat{\delta^{(s)}_n}\!\!=\!\!\widehat{A^n\delta^{(s)}_0}\!{}&=\!\Lambda^n \cdot \frac{1}{\sqrt{N}}1\!\!=\!\!\frac{1}{\sqrt{N}}\lambda^n
\Longrightarrow \delta^{(s)}_n\!\!=\!\!\frac{1}{\sqrt{N}}\textrm{GFT}^{-1}\!\!\left(\lambda^n\right)
\\
\label{eqn:GSPimpulseshiftdeltas-5}
{}&\textrm{GFT}^{-1}\left(M^n\widehat{\delta^{(s)}_0}\right)= \Lambda^{*^n}\delta^{(s)}_0=\frac{1}{\sqrt{N}}\sum_{i=0}^{N-1}\lambda^n\odot v_i
\end{align}
Define the impulse matrix $D^{(s)}$ of $\delta^{(s)}_0$ and its shifts like in~\eqref{eqn:matriximpv}; its $\textrm{GFT}$ is the normalized Vandermonde (no factor $\mathcal{Y}_0$)
\begin{align}
\label{eqn:GSPimpulseshiftdeltas-6}
\widehat{D^{(s)}}{}&=\mathcal{V}_{\lambda}
\end{align}
For this definition of the graph impulse, we only need assumption~\ref{assp:diagonalizableA} for~$D^{(s)}$ and $\widehat{D^{(s)}}$ to be invertible.

 The two alternatives lead to different tradeoffs and the results are consistent with shifting delta functions in DSP.
	\section{Modulation, Filtering, and Convolution}\label{sec:modconvolution}
In DSP \cite{oppenheimwillsky-1983,siebert-1986}, modulation is the entry wise (or Hadamard) product of signals and occurs in communications systems like in (amplitude) modulation where a (message) signal modulates a carrier. Filtering is widely used to reduce noise or shape signals. Filtering and convolution are tightly connected as we discuss here. Further, it is well known that product (modulation) in one domain corresponds to convolution (or filtering) in the other domain. References \cite{Sandryhaila:13,Sandryhaila:14} study GSP filtering in the vertex domain defined by a matrix-vector multiplication. For shift invariant filters, the filters are polynomial on the shift as in~\eqref{convtime}. References~\cite{Sandryhaila:13,Sandryhaila:14} also show that filtering in the spectral domain is the pointwise product (modulation) of the GFT of the input signal with the graph frequency response of the polynomial filter. This section completes the picture for GSP with respect to modulation, filtering, and convolution of graph signals in both the vertex and graph spectral domains, showing the relation among these concepts. It shows how to design polynomial filters given their graph spectral response, graph impulse response, how to convolve explicitly two graph signals, and introduces polynomial spectral filters $P(M)$.
\vspace*{-.4cm}
	\subsection{Filtering in Vertex- and Modulation in Spectral-Domains} \label{subsec:filteringVmodS}
	Consider the polynomial filter in the shift~$A$
	\begin{align}\label{PAeq}
	P(A) {}&= p_{0}I+p_{1}A^{1}+\hdots+p_{N-2}A^{N-2}+p_{N-1}A^{N-1}\\
    \label{Apoly}
    {}&=\textrm{GFT}^{-1}P(\Lambda)\textrm{GFT}.
	\end{align}
	and let~$x$ be the input graph signal to $P(A)$. In the graph spectral domain, filtering is modulation (pointwise multiplication) of~$\widehat{x}$ by the graph frequency response of the filter~$\widehat{y}$ \cite{Sandryhaila:13,Sandryhaila:14}
\vspace*{-.2cm}
	\begin{align}\label{convpairtime}
	\hspace{-.6cm}(\textrm{\color{blue}{vertex filtering}})\:\:P(A) \cdot x{}& \xrightarrow{\mathcal{F}} \widehat{y} \odot \widehat{x}\:\:\:(\textrm{\color{red}{spectral modulation}})\\
    \label{eqn:convpairtime-1}
    \hspace{-.6cm}(\textrm{\color{red}{spectral response}})\:\:\: \widehat{y}&\triangleq P\left(\Lambda\right)\cdot 1 \textrm{   and   }
	P(\Lambda) = \textrm{diag}\left[\widehat{y}\right]\\
	\label{result1}
    P(\Lambda)\cdot\widehat{x} {}&= \widehat{y} \odot \widehat{x}\\
    \label{convtimeans}
	P(A){}&= \textrm{GFT}^{-1} \cdot \textrm{diag}\left[\widehat{y}\right]\cdot \textrm{GFT}
	\end{align}
	where~$\odot$ is the Hadamard product, $\triangleq$ defines the quantity on the left, and~$1$ is the vector of ones. From~\eqref{convpairtime}, modulation in the spectral domain of $\widehat{y}$ and $\widehat{x}$  is filtering of~$x$ by the graph filter $P(A)=\textrm{GFT}^{-1} \textrm{diag}\left[\widehat{y}\right] \textrm{GFT}$  with frequency response $\widehat{y}$.
	\vspace*{-.3cm}
	\subsection{Filtering in Spectral- and Modulation in Vertex-Domains}
	With~$M$ in~\eqref{eqn:Mgeqn}, let the spectral polynomial filter
\begin{align}\label{eqn:PMeq}
    P(M) {}&= p_{0}I+p_{1}M^{1}+\hdots+p_{N-2}M^{N-2}+p_{N-1}M^{N-1}\\
    \label{Apolyf}
{}&=\textrm{GFT}\cdot P\left(\Lambda^*\right)\cdot \textrm{GFT}^{-1}.
	\end{align}
Spectral filtering of the input~$\widehat{x}$ by $P(M)$ is
\vspace*{-.2cm}
	\begin{align}\label{convpair}
	\hspace{-.6cm}(\textrm{\color{red}{spectral filtering}})\:\:P(M) \cdot \widehat{x}  {}&\xrightarrow{\mathcal{F}^{-1}} y \odot x \:(\textrm{\color{blue}{vertex modulation}})\\
    \label{convpair-1}
    \hspace{-.6cm}(\textrm{\color{blue}{`vertex response'}})\:\:y{}&=P\left(\Lambda^*\right)\cdot 1 \textrm{   and   }
	P(\Lambda^*) = \textrm{diag}[y] \\
    \label{result2}
	P(\Lambda^*) \cdot x {}&= y \odot x\\
    \label{convfreqans}
	P(M) {}&= \textrm{GFT} \cdot \textrm{diag}[y] \cdot \textrm{GFT}^{-1}
	\end{align}
	Thus, filtering of~$\widehat{x}$ in the graph spectral domain by $P(M)$ is modulation in the vertex domain of~$x$ by $y=P\left(\Lambda^*\right)\cdot 1$.
\vspace*{-.3cm}
\subsection{Convolution and Filtering in the Vertex Domain}\label{subsec:convolutionV}
In \cite{Sandryhaila:13,Sandryhaila:14big}, convolution in the vertex domain is filtering the graph signal by a graph polynomial filter $P(A)$. But these references nor any other available address the convolution of two graph signals~$x$ and~$y$ when the graph filter $P(A)$ is not known. To convolve~$x$ with~$y$, let~$x$ be input to~$P(A)$ and~$y$ be its graph impulse response.  Then, in the vertex domain
\vspace*{-.1cm}
\begin{align}
\label{eqn:gspconv-poly}
\hspace{-.5cm}(\textrm{\color{red}{vertex convolution}})\:\:\:y\star x{}&=P(A)\cdot x \:\:\:(\textrm{\color{blue}{vertex filtering}})
\end{align}
\vspace*{-.1cm}
To solve~\eqref{eqn:gspconv-poly}, consider how to determine $P(A)$ from its impulse response~$y$. Let $P(A)$ be given in~\eqref{PAeq}. The graph impulse response~$y$ is the response of the filter $P(A)$ to the graph impulse $\delta_0$. Using~\eqref{PAeq} in~\eqref{eqn:gspconv-poly} and either definition of the graph impulse and its shifts in section~\ref{sec:deltafunction} (omitting the superindex $(v)$ or $(s)$), get successively
\vspace*{-.2cm}
\begin{align}
\label{eqn:gspconv-poly-2}
\hspace{-.5cm}(\textrm{\color{blue}{graph impulse response}})\:\:\:\:\:P(A)\cdot \delta_0{}&=y\\
\label{eqn:gspconv-poly-3}
p_0 \delta_0+p_1A\cdot\delta_0+\cdots+p_{N-1} A^{N-1}\delta_0{}&=y\\
\label{eqn:gspconv-poly-4}
p_0 \delta_0+p_1\delta_1+\cdots+p_{N-1} \delta_{N-1}{}&=y\\
\label{eqn:gspconv-poly-5}
D\cdot p{}&=y
\end{align}
where~$D$ is the impulse matrix collecting the impulse and its shifts (e.g., \eqref{eqn:matriximpv}) and the unknown vector of the polynomial filter coefficients is $p=\left[p_0\cdots p_{N-1}\right]^T$. If~$D$ is invertible, solving~\eqref{eqn:gspconv-poly-5} for~$p$ by any available method, for example, Gauss elimination, gives~$p$ and the graph filter polynomial $P(A)$ that defines the convolution~\eqref{eqn:gspconv-poly} of the two graph signals~$x$ and~$y$. If we adopt $\delta^{(v)}$, invertibility of~$D$ needs both Assumptions~\ref{assp:diagonalizableA} and~\ref{assp:nonzeroy0}; if instead we work with $\delta^{(s)}$, invertibility of~$D$ requires only Assumption~\ref{assp:diagonalizableA}. Note that solving~\eqref{eqn:gspconv-poly-5} does not require the graph spectral factorization of~$A$ nor direct verification of Assumptions~\ref{assp:diagonalizableA} or~\ref{assp:nonzeroy0} since using~\eqref{eqn:matriximpv-6} the impulse matrix~$D$ is obtained directly in the vertex domain in terms of $a_o^{(n)}$, $n=0\cdots N-1$, and its rank found by any appropriate numerical method (again, for example by Gauss elimination).

In alternative, take the $\textrm{GFT}$ of both sides of~\eqref{eqn:gspconv-poly-5} to get
\begin{align}
\label{eqn:gftsolnP}
\mathcal{V}_{\lambda}\cdot p{}&=\widehat{y}
\end{align}
Under assumption~\ref{assp:diagonalizableA} only, or under  assumptions~\ref{assp:diagonalizableA} and~\ref{assp:nonzeroy0}, $\mathcal{V}_{\lambda}$ is full rank and~\eqref{eqn:gftsolnP} determines uniquely~$p$ and $P(A)$.
\vspace*{-.2cm}
\begin{remark}\label{rem:soltnfilter}
For large~$N$, it is not practical to solve either~\eqref{eqn:gspconv-poly-5} or~\eqref{eqn:gftsolnP}. Resorting to low rank sparse solutions solve either
\vspace*{-.2cm}
\begin{align}
\label{eqn:sparsifiedslnP-1}
\widehat{p}{}&=\arg\min_z\left\{\left\|y-D\cdot z\right\|^2_2+\left|z\right|_1\right\}\\
\label{eqn:sparsifiedslnP-2}
\widehat{p}{}&=\arg\min_z\left\{\left\|\widehat{y}-\mathcal{V}_{\lambda}\cdot z\right\|^2_2+\left|z\right|_1\right\}
\end{align}
\end{remark}
\vspace*{-.6cm}
\subsection{Convolution and Filtering in the Spectral Domain}\label{subsec:convolutionS}
We now consider convolution of two graph signals $\widehat{x}$ and~$\widehat{y}$ through filtering in the spectral domain by a spectral graph polynomial filter $P(M)$ given in~\eqref{eqn:PMeq} where~$M$ is the spectral shift, see~\eqref{eqn:Mgeqn}. We follow section~\ref{subsec:convolutionV}. In the spectral domain
\begin{align}
\label{eqn:gspconv-polyS}
\hspace{-.5cm}(\textrm{\color{red}{spectral convolution}})\:\:\:\widehat{y}\star \widehat{x}{}&=P(M)\cdot \widehat{x} \:\:\:(\textrm{\color{blue}{spectral filtering}})
\end{align}
To move further and following the recipe in equations~\eqref{eqn:gspconv-poly-2} through~\eqref{eqn:gspconv-poly-5}, we first introduce an impulse in the spectral domain $\widehat{\delta_0^{\scriptsize\textrm{sp}}}$. Like in section~\ref{sec:deltafunction}, we can have two versions, but now $\widehat{\delta_0^{\scriptsize\textrm{sp}(s)}}=e_0$ is impulsive and $\delta_0^{\scriptsize\textrm{sp}(v)}=\frac{1}{\sqrt{N}}1$ is flat. To convolve in the spectral domain, solve for the vector~$p$ of coefficients of $P(M)$ by taking $\widehat{y}$ as its impulse response
\begin{align}
\label{eqn:gspconv-polyS-2}
\hspace{-.9cm}(\textrm{\color{red}{spectral filter graph impulse response}})\:\:\:P(M)\cdot \widehat{\delta_0^{\scriptsize\textrm{sp}}}{}&=\widehat{y}\\
\label{eqn:gspconv-polyS-3}
p_0 \widehat{\delta_0^{\scriptsize\textrm{sp}}}{}+p_1M\cdot \widehat{\delta_0^{\scriptsize\textrm{sp}}}{} +\cdots+p_{N-1} M^{N-1}\widehat{\delta_0^{\scriptsize\textrm{sp}}}{}{}&=\widehat{y}\\
\label{eqn:gspconv-polyS-4}
p_0 \widehat{\delta_0^{\scriptsize\textrm{sp}}}{}+ p_1\widehat{\delta_1^{\scriptsize\textrm{sp}}}{} +\cdots+p_{N-1} \widehat{\delta_{N-1}^{\scriptsize\textrm{sp}}}{}{}&=\widehat{y}\\
\label{eqn:gspconv-polyS-5}
E\cdot p{}&=\widehat{y}
\end{align}
where the matrix~$E$ collects the impulse $\widehat{\delta_0^{\scriptsize\textrm{sp}}}{}$ and its shifts.
\vspace{-.2cm}	
	\section{GSP Sampling} \label{sec:gspsampling}
    We consider two main problems in GSP Sampling Theory:
    \begin{enumerate}
        \item How to choose a sampling set~$\mathcal{S}$;
        \item Given a sampled signal $x_s$, recover the original signal~$x$.
    \end{enumerate}
    We present from DSP first principles GSP solutions in the vertex and spectral domains and  relations between the two.
\vspace*{-.5cm}
    \subsection{GSP Sampling in the Vertex Domain\textemdash Sampling Set} \label{subsec:samplingset}
	Assume the graph signals~$x\in\mathcal{X}$ are bandlimited with bandwidth~$K$ \cite{pesenson2008sampling,Jelena,anis2016efficient}. A sampling set~$\mathcal{S}$ \cite{pesenson2008sampling} for signals~$x$ is a subset of graph nodes or vertices such that the original signal~$x$ is perfectly reconstructed from its samples indexed by the nodes in~$\mathcal{S}$. We define~$\mathcal{S}$ by its indicator function $\delta^{\scriptsize\textrm{(spl)}}$ that we refer to as the sampling signal\textemdash $\delta^{\scriptsize\textrm{(spl)}}_i=1, \textrm{ if } i\in\mathcal{S}$, and 0 otherwise. We assume that the graph frequencies have been ordered \cite{Sandryhaila:14,Sandryhaila:14big} and without loss of generality (wlog) that $\widehat{x}$ is lowpass and bandlimited to bandwidth~$K$. In DSP, the sampling signal $\delta^{\scriptsize\textrm{(spl)}}$ is a train of \textit{equally spaced} time delta impulses and sampling is \textit{modulation} (product) of~$x$ by $\delta^{\scriptsize\textrm{(spl)}}$, i.e., the sampled signal (before decimation) is $x\odot \delta^{\scriptsize\textrm{(spl)}}$ \cite{oppenheimwillsky-1983,siebert-1986}.
	
	We now find the sampling set~$\mathcal{S}$ by finding its indicator $\delta^{\scriptsize\textrm{(spl)}}$. Since~$x$ is lowpass, we split $\widehat{x}$ into two parts: $\widehat{x}_{K}$ is the first $K$ entries of $\widehat{x}$ (containing all the non-zero entries) and $\mathbf{0}_{N-K}$ the remaining $N-K$ $0$s after the first~$K$ entries:
	\begin{align} \label{xhatg}
	\widehat{x} {}&= \left[\begin{array}{c}
\widehat{x}_{K}\\
\textbf{0}
\end{array}\right]
	\end{align}
	Split $\textrm{GFT}$ into its first~$K$ rows and remaining~$N-K$ rows:
	\begin{align} \label{gftsplit}
	\textrm{GFT} = \begin{bmatrix}
	\textrm{GFT}_{K} \\ \textrm{GFT}_{N-K}
	\end{bmatrix}
	\end{align}
Recall digraph $G=(V,E)$, $V$ set of nodes, $E$ set of edges.
\begin{theorem}\label{thm:samplingsetGE1}
If the signal bandwidth is~$K=N$, then $\mathcal{S}=V$. Otherwise, a (not necessarily unique) sampling set~$\mathcal{S}$ is given by a set of~$K$ free variables in the solution of
	\begin{align}
	\textrm{GFT}_{N-K}\cdot x &= \textbf{0}\label{split-0}
	\end{align}
\end{theorem}
\begin{proof}
    We start by observing that with a linear system $H\cdot x=b$ of~$N$ equations in~$N$ unknowns or variables~$x$, the solution exists and is unique if and only if~(iff) $H$ is full rank. We can
    \begin{inparaenum}[1)]
    \item choose any subset of $N-K$ equations in the~$N$ unknowns~$x$,
     \item express $N-K$ of the unknowns in terms of the other~$K$ (even if trivially), \item replace these $N-K$ expressions in the remaining~$K$ equations, and
            \item solve these~$K$ equations for these~$K$ unknowns.
                \end{inparaenum}
         In the sequel, we call the $N-K$ unknowns as \textit{pivots} variables and the~$K$ unknowns as \textit{free} variables.

    We consider first $K<N$. Since the GFT is invertible, $\textrm{GFT}\cdot x=\widehat{x}$ is a full rank linear system. From~\eqref{xhatg} and~\eqref{gftsplit}, get
	\begin{align}
    \label{split-1}
	\textrm{GFT}_{K}\cdot x &= \widehat{x}_K\\
	\textrm{GFT}_{N-K}\cdot x &= \textbf{0}_{N-K}
    \label{split}
	\end{align}
	 Since $\textrm{rank}(\textrm{GFT}_{N-K}) = N-K$, the $N-K$ equations~\eqref{split} in~$N$ unknowns are linearly independent~(l.i.); we apply Gauss-Jordan elimination to row reduce the $(N-K) \times N$ matrix, $\textrm{GFT}_{N-K}$. This yields $N - K$ pivot variables and~$K$ free variables. To recover the original signal~$x$, we only need the~$K$ entries of~$x$ that are the free variables. The other $N-K$ pivot variables are determined from the values of the~$K$ free variables. Thus, the sampling set~$\mathcal{S}$ is the set of nodes indexing the free variables. Its indicator function $\delta^{\scriptsize\textrm{(spl)}}$ is:
	\begin{align} \label{deltaif}
	 \delta^{\scriptsize\textrm{(spl)}} =
  \begin{cases}
   1, & \text{at each free variable location}\\
    0, & \text{at each pivot variable location}\\
  \end{cases}
  \end{align}
  Because there is freedom in choosing the free variables, $\mathcal{S}$ is not unique.
  If $K=N$, then $N-K=0$ and~\eqref{split} is the trivial equation, there are no pivots and~$\mathcal{S}=V$.
  \end{proof}
\vspace*{-.5cm}
   \subsection{GSP Sampling in the Vertex Domain\textemdash Recovery} \label{subsec:samplingvertex}
    By Gauss elimination of $\textrm{GFT}_{N-K}$, we obtain a sampling set~$\mathcal{S}$, $\delta^{\scriptsize\textrm{(spl)}}$ in~\eqref{deltaif},  $K$ free variables, and $N-K$ pivot variables. Let the free and pivot variables be $x_f=\left[x_{f_1},x_{f_2},\cdots,x_{f_K}\right]^T$ and $x_p=\left[x_{p_1},x_{p_2},\cdots,x_{p_{N-K}}\right]^T$. From Gauss elimination, every pivot is a linear combination of free variables.
%
%
    In matrix form:
    \vspace*{-.2cm}
    \begin{align}\label{pivottofree}
     \underbrace{\begin{bmatrix}
	   x_{p_1}\\
	   x_{p_2}\\
	   \vdots\\
	   x_{p_{N-K}}
	\end{bmatrix}}_{x_p}
  &=  \underbrace{\begin{bmatrix}
	    s_{1,1} & s_{1,2}  &\hdots  &s_{1,K} \\
	    s_{2,1} & s_{2,2}  &\hdots  &s_{2,K} \\
	    \vdots  &\vdots & \ddots  &\vdots \\
	    s_{N-K,1}  &s_{N-K,2} & \hdots  &s_{N-K,K}
	\end{bmatrix}}_{S}
	 \underbrace{\begin{bmatrix}
	   x_{f_1}\\
	   x_{f_2}\\
	   \vdots\\
	   x_{f_K}
	\end{bmatrix}}_{x_f}
  \end{align}
  Let $x_s = x_f = [x_{f_1},x_{f_2},\cdots,x_{f_K}]^T$ be the sampled and decimated signal in the vertex domain. To recover the original~$x$ from~$x_s$, first get $x_p$ from~\eqref{pivottofree} and then recover~$x$ by reordering the entries of~$x_s$ and~$x_p$ into their correct locations.
\vspace*{-.2cm}
\begin{algorithm}
{\small \tt
\caption{\label{alg:algorithm1}\scriptsize GSP Sampling: Vertex Domain Recovery}
\textbf{Given:} Sampled/sampling signals~$x_s$ and $\delta^{\scriptsize\textrm{(spl)}}$\\
By Gauss Elimination of \eqref{split}, form $S$ in \eqref{pivottofree} \\
Use \eqref{pivottofree} and $x_f = x_s$ to find $x_p$\\
Reorder/ combine $x_f$ and $x_p$ entries to get~$x$
}
\end{algorithm}
\vspace*{-.9cm}
   \subsection{GSP Sampling in the Spectral Domain\textemdash Overview} \label{subsec:samplingspectral}
	Consider an arbitrary graph~$G$ and its adjacency matrix~$A$. Let~$x$ be a bandlimited graph signal, $\left\|\widehat{x}\right\|_0 \leq K$, and $\delta^{\scriptsize\textrm{(spl)}}$ the sampling signal, $\left\|\delta^{\scriptsize\textrm{(spl)}}\right\|_0 = K$. As noted before, sampling~$x$ in the vertex domain is modulation $\delta^{\scriptsize\textrm{(spl)}}$, $x\odot \delta^{\scriptsize\textrm{(spl)}}$. By~\eqref{convpair}, this is equivalent to filtering in the spectral graph domain:
	\begin{align} \label{gspsampling}
	\delta^{\scriptsize\textrm{(spl)}} \odot x  \xrightarrow{\mathcal{F}} P(M) \cdot \widehat{x}
	\end{align}
	In the frequency domain, sampling is the matrix vector product $P(M)\cdot\widehat{x}$. Our approach to recover the signal~$x$ from its sampled and decimated version $x_s$ is based on interpretation~\eqref{gspsampling} of sampling in the vertex domain as spectral filtering by a polynomial $P(M)$ in the spectral shift~$M$.
	We start by determining $P(M)$. Using~\eqref{convfreqans} with $y=\delta^{\scriptsize\textrm{(spl)}}$, $P(M)$ is:
\vspace*{-.1cm}
	\begin{align} \label{gspsamp}
	P(M) &= \textrm{GFT} \ \textrm{diag}\left[\delta^{\scriptsize\textrm{(spl)}}\right] \ \textrm{GFT}^{-1} \\
\label{gspsamp-2}
	{}&=  \textrm{GFT} \ \textrm{diag}\left[\delta^{\scriptsize\textrm{(spl)}}\right]\ \textrm{diag}\left[\delta^{\scriptsize\textrm{(spl)}}\right] \ \textrm{GFT}^{-1}
	\end{align}
	Equation~\eqref{gspsamp-2} follows from~\eqref{gspsamp} since $\delta^{\scriptsize\textrm{(spl)}}$ contains only ones and zeros. The matrix $\textrm{diag}\left[\delta^{\scriptsize\textrm{(spl)}}\right]$ in
	$\textrm{GFT}\ \textrm{diag}\left[\delta^{\scriptsize\textrm{(spl)}}\right]$ chooses columns of the GFT and in $\textrm{diag}\left[\delta^{\scriptsize\textrm{(spl)}}\right] \ \textrm{GFT}^{-1}$ chooses rows of $\textrm{GFT}^{-1}$. If column~$i$ is chosen from the GFT, then row~$i$ is chosen from $\textrm{GFT}^{-1}$.
	Since $\left\|\widehat{x}\right\|_0 \leq K$, we assume wlog that the first~$K$ entries of~$\widehat{x}$ contain its non-zero entries and split~$\widehat{x}$ as in~\eqref{xhatg} into two parts: $\widehat{x_{K}}$ with the first~$K$ entries of~$\widehat{x}$ (that contain all its non-zero entries)\footnote{\label{ftn:split-1}Assumption~\eqref{xhatg} that the ``band'' occurs in the first~$K$ entries is not necessary. We can let $\widehat{x}_{K}$  contain all the non-zero entries, regardless of where they are in $\widehat{x}$. Then, $\textrm{GFT}^{-1}_{K}$ refers to the~$K$ columns corresponding to the~$K$ chosen entries of $\widehat{x}$ instead of its first~$K$ columns. To simplify the notation, we assumed in~\eqref{xhatg} that the ``band'' occurs in the first~$K$ entries. If not, by permuting rows and columns of the matrices, partition~\eqref{xhatg} is always possible.} and $\textbf{0}_{N-K}$ the remaining $N-K$ $0$s. We also partition $\textrm{GFT}$,  $\textrm{GFT}^{-1}$, and $P(M)$ as\footnote{We emphasize that these partitions in~\eqref{eqn:partitionGFTcl-1} and~\eqref{eqn:partitionGFTcl-2} are now columnwise, in contrast with the partition of the $\textrm{GFT}$ in~\eqref{gftsplit} in section~\ref{subsec:samplingset} that was row-wise. Hopefully this will not distract or confuse the reader.}
	\begin{align}
    \label{eqn:partitionGFTcl-1}
	\textrm{GFT}\phantom{^{-1}} {}&= \begin{bmatrix}
	\textrm{GFT}_{K} &\hspace{+.1cm} \textrm{GFT}_{N-K}
	\end{bmatrix}\\
    \label{eqn:partitionGFTcl-2}
	\textrm{GFT}^{-1} {}&= \begin{bmatrix}
	\textrm{GFT}^{-1}_{K} & \hspace{-.04cm}\textrm{GFT}^{-1}_{N-K}
	\end{bmatrix}\\
    \label{eqn:partitionGFTcl-3}
	P(M) {}&= \begin{bmatrix}
	P(M)_{K} &\hspace{-.15cm} P(M)_{N-K}
	\end{bmatrix}
	\end{align}
	Using~\eqref{xhatg}, we can write:
	\begin{align} \label{pmhat-0}
	P(M)\cdot\widehat{x}{}&=P(M)_K\cdot\widehat{x}_K{}\\
    \label{pmhat}
    {}&=\textrm{GFT} \, \textrm{diag}\left[\delta^{\scriptsize\textrm{(spl)}}\right] \! \textrm{diag}\left[\delta^{\scriptsize\textrm{(spl)}}\right]\! \textrm{GFT}^{-1}_{K} \cdot \widehat{x}_{K}
	\end{align}
	The block $P(M)_K=\textrm{GFT} \ \textrm{diag}\left[\delta^{\scriptsize\textrm{(spl)}}\right]\ \textrm{diag}\left[\delta^{\scriptsize\textrm{(spl)}}\right] \ \textrm{GFT}^{-1}_{K}$ is $N\times K$. To recover~$x$, we need first to recover $\widehat{x}_K$. We can achieve this if we find~$K$ l.i.~rows of $P(M)_K$ to form an invertible $K\times K$ matrix, and then recover $\widehat{x}_K$ by inverting this $K\times K$ block of $P(M)_K$. The next sections detail this.
\vspace*{-.6cm}
   \subsection{GSP Sampling in the Spectral Domain\textemdash Sampling Set} \label{subsec:samplingsetspectral}
    Again, the sampling signal $\delta^{\scriptsize\textrm{(spl)}}$ is the indicator function of the sampling set~$\mathcal{S}$, and our goal is to determine it such that the bandlimited graph signal~$x$ is recoverable from a sampled version~$x_s$.
   From~\eqref{pmhat}, for~$x$ to be recoverable, $\delta^{\scriptsize\textrm{(spl)}}$ must be chosen so that the $N\times K$ matrix $P(M)_K=\textrm{GFT} \ \textrm{diag}\left[\delta^{\scriptsize\textrm{(spl)}}\right]\ \textrm{diag}\left[\delta^{\scriptsize\textrm{(spl)}}\right] \ \textrm{GFT}^{-1}_{K}$ in~\eqref{pmhat} contains~$K$ l.i.~rows.
    This problem can be solved by considering all possible choices for $\delta^{\scriptsize\textrm{(spl)}}$ in order to find the one that leads to~$K$ l.i.~rows. This has combinatorial complexity and is not feasible. Instead, we consider an alternative to design $\delta^{\scriptsize\textrm{(spl)}}$.

    We define the sampling signal $\delta^{\scriptsize\textrm{(spl)}}$ by choosing~$K$ l.i.~rows of the $N\times K$ matrix $\textrm{GFT}^{-1}_{K}$.
    \begin{align} \label{deltaiffreq}
	 \delta^{\scriptsize\textrm{(spl)}} =
  \begin{cases}
   1, & \text{if one of the chosen $K$ l.i.~rows}\\
    0, & \text{if not one of the chosen $K$ l.i.~rows}\\
  \end{cases}
  \end{align}
   \begin{theorem} \label{deltaexists}
   The $\delta^{\scriptsize\textrm{(spl)}}$ defined in Equation \eqref{deltaiffreq} always exists.
   \end{theorem}
  \begin{proof}
  $\textrm{GFT}^{-1}$ is full rank. Then, its first~$K$ columns are l.i.~and $\textrm{rank}(\textrm{GFT}^{-1}_{K}) = K$. Thus, $\textrm{GFT}^{-1}_{K}$ has $K$ l.i.~rows.
  \end{proof}

  Using this choice of $\delta^{\scriptsize\textrm{(spl)}}$, we have the following Theorem:
   \begin{theorem} \label{deltatheorem}
	The $N\times K$ matrix
\begin{align}
\label{eqn:defineL}
P(M)_K{}&=\textrm{GFT} \ \textrm{diag}\left[\delta^{\scriptsize\textrm{(spl)}}\right]\ \textrm{diag}\left[\delta^{\scriptsize\textrm{(spl)}}\right] \ \textrm{GFT}^{-1}_{K}
\end{align}
contains~$K$  linearly independent~(l.i.) rows if and only if the $N\times K$ matrix $\textrm{GFT}^{-1}_{K}$ contains~$K$ linearly independent rows.
   \end{theorem}
   \begin{proof}
\textit{If}. Let $R = \textrm{GFT}^{-1}_{K}$. By assumption the $N\times K$ matrix~$R$ has~$K$ l.i.~rows; wlog, assume they are the top~$K$ rows of~$R$ (by~\eqref{xhatg} they are, obtained by row and column matrix reordering). With the $\textrm{GFT}$ columnwise split in~\eqref{eqn:partitionGFTcl-1}, let~$R_K$ be the $K\times K$ block of the first l.i.~$K$ rows of~$R$ and $R_{N-K}$ the $N-K\times K$ block with the remaining $N-K$ rows. Then
	\begin{align}\label{eqn:blockL}
		P(M)_K{}&\!\!\!=\!\!\!\begin{bmatrix}
	\textrm{GFT}_K \, \textrm{GFT}_{N-K}
	\end{bmatrix}
    \!\!\textrm{diag}\!\!\left[\delta^{\scriptsize\textrm{(spl)}}\right] \!\!\textrm{diag}\!\!\left[\delta^{\scriptsize\textrm{(spl)}}\right]
	\!\!\!
    \begin{bmatrix}
	R_{K} \\ R_{N-K}
	\end{bmatrix}\,
\end{align}
	In~\eqref{eqn:blockL}, $\delta^{\scriptsize\textrm{(spl)}}$ chooses the~$K$ l.i.~rows of~$R$ and zeros out the others. It also chooses the corresponding~$K$ columns of the $\textrm{GFT}$ and zeros out the remaining columns. In other words,
	\begin{align}\label{GFTReq-1}
	P(M)_K{}&=\begin{bmatrix}
	\textrm{GFT}_K & \textbf{0}
	\end{bmatrix}
	\begin{bmatrix}
	R_{K} \\ \textbf{0}
	\end{bmatrix}
=\textrm{GFT}_K R_{K}
\end{align}
Since $\textrm{GFT}$ is invertible, it is full rank and all its $N$ columns are l.i. Since $\textrm{GFT}_K$ is formed from~$K$ columns of $\textrm{GFT}$, $\textrm{rank}(\textrm{GFT}_K) = K$, and~$K$ of its~$N$ rows are l.i. Let $\textrm{GFT}_{K_K}$ be the $K\times K$ block of these~$K$ l.i.~rows of $\textrm{GFT}_K$. Let
\begin{align} \label{GFTRL}
   P(M)_{K_K}{}&= \textrm{GFT}_{K_K} R_{K}
\end{align}
 Since $P(M)_{K_K}$ is the product of two full rank $K\times K$ matrices, it is full rank and invertible. Since $P(M)_{K_K}$ is formed from the rows of~$P(M)_K$ in~\eqref{GFTReq-1}, $P(M)_K$ has~$K$ l.i.~rows.

\textit{Only if}. Taken literally, the converse needs no proof because $\textrm{GFT}^{-1}_K$ has rank~$K$ since it is a $N\times K$ block of the full rank $\textrm{GFT}^{-1}$. What needs to be proven is the following. By assumption, the $N\times K$ matrix $P(M)_K$ with rank~$K$ has a decomposition like in~\eqref{eqn:defineL}, rewritten as
\begin{align}
\label{eqn:defineL-1}
P(M)_K{}&=F_1\,\,\textrm{diag}\left[\delta^{\scriptsize\textrm{(spl)}}\right]\ \textrm{diag}\left[\delta^{\scriptsize\textrm{(spl)}}\right] \ F_2
\end{align}
where $F_1$ is $N\times N$, $F_2$ is $N\times K$, and $\textrm{diag}\left[\delta^{\scriptsize\textrm{(spl)}}\right]$ is a diagonal matrix with~$K$ nonzero entries that are all equal to one. The positions of the nonzero diagonal entries of $\textrm{diag}\left[\delta^{\scriptsize\textrm{(spl)}}\right]$ correspond to~$K$ rows of $P(M)_K$ that are linearly independent\footnote{We know that~$K$ of the~$N$ rows of $P(M)_K$ are l.i., but this set is not unique.}. Given this set-up, what needs to be proven is that $F_2$ is full rank and so it has~$K$ l.i.~rows. In the sequel, for simplicity, we work with~\eqref{eqn:defineL}, and so consider $F_1=\textrm{GFT}$ and $F_2=\textrm{GFT}^{-1}_K$, but still look to prove directly that $\textrm{GFT}^{-1}_K$ has rank~$K$.

Assume $\textrm{rank}(P(M)_K)= K$ and so with~$K$ l.i.~rows. Wlog, assume these are the first~$K$ rows of~$P(M)_K$. Split~$P(M)_K$ rowwise with the $K\times K$ block $P(M)_{K_K}$ with its first~$K$ rows and the $N-K\times K$ block $P(M)_{K_{N-K}}$ with the remaining $N-K$ rows of~$P(M)_K$. Similarly, split the $\textrm{GFT}$ matrix rowwise with the $K\times N$ block $\textrm{GFT}_{K}$ with its first~$K$ rows and the $N-K\times N$ block $\textrm{GFT}_{N-K}$ with the remaining $N-K$ rows\footnote{We emphasize that in the ``Only if'' part of the proof, we split the $\textrm{GFT}$ rowwise~\eqref{gftsplit} rather than columnwise~\eqref{eqn:partitionGFTcl-1} as assumed in the ``If'' part.}.  This yields:
\begin{align}
    \label{expandedgft-1}
	P(M)_K{}&=\begin{bmatrix}
	P(M)_{K_{K}} \\ P(M)_{K_{N-K}}
	\end{bmatrix}\\
    {}&=\begin{bmatrix}
    \textrm{GFT}_{K} \\ \textrm{GFT}_{N-K}
	\end{bmatrix} \textrm{diag}\left[\delta^{\scriptsize\textrm{(spl)}}\right] \textrm{diag}\left[\delta^{\scriptsize\textrm{(spl)}}\right]
	\textrm{GFT}^{-1}_{K}\\
	 \label{expandedgft}
	P(M)_{K_{K}}{}&= \textrm{GFT}_{K}\,\,\textrm{diag}\left[\delta^{\scriptsize\textrm{(spl)}}\right] \textrm{diag}\left[\delta^{\scriptsize\textrm{(spl)}}\right]
	\textrm{GFT}^{-1}_{K}
\end{align}
From~\eqref{deltaiffreq}, in $\textrm{diag}\left[\delta^{\scriptsize\textrm{(spl)}}\right]\textrm{GFT}^{-1}_{K}$, $\textrm{diag}\left[\delta^{\scriptsize\textrm{(spl)}}\right]$ zeros out (the bottom) $N-K$ rows of
	$\textrm{GFT}^{-1}_{K}$. Denote the remaining~$K$ rows as $\textrm{GFT}^{-1}_{K\delta^{\scriptsize\textrm{(spl)}}_{i}}$, $i = 0\cdots K-1$. Similarly, in $\textrm{GFT}_{K}\,\,\textrm{diag} \left[\delta^{\scriptsize\textrm{(spl)}}\right]$, $\textrm{diag}\left[\delta^{\scriptsize\textrm{(spl)}}\right]$ zeros out the (most right) corresponding $N-K$ columns. Denote the remaining~$K$ columns as $\textrm{GFT}_{K\delta^{\scriptsize\textrm{(spl)}}_{i}}$, $i = 0\cdots K-1$. Rewrite~\eqref{expandedgft} as:
	\begin{align} \label{deltagone}
	P(M)_{K_K}{}&\!\!\!=\!\!\!\begin{bmatrix}
	\textbf{0}& \!\!\!\!\!\textrm{GFT}_{K\delta^{\scriptsize\textrm{(spl)}}_{0}}& \!\!\!\!\!\textbf{0}&\!\!\!\!\! \vdots & \!\!\!\!\!\textbf{0}& \!\!\!\!\!\textrm{GFT}_{K\delta^{\scriptsize\textrm{(spl)}}_{K-1}}&\!\!\!\!\!\!\!\textbf{0}
	\end{bmatrix}
\!\!\!\!
	\left[\!\!\!\begin{array}{ccccccc}
	\textbf{0} \\ \textrm{GFT}^{-1}_{K\delta^{\scriptsize\textrm{(spl)}}_{0}} \\ \textbf{0} \\ \vdots \\ \textbf{0} \\ \textrm{GFT}^{-1}_{K\delta^{\scriptsize\textrm{(spl)}}_{K-1}} \\ \textbf{0}
	\end{array}\!\!\!\right]\hspace{.26cm}
	\end{align}
where $\textbf{0}$ represents the entries where $\delta^{\scriptsize\textrm{(spl)}} = 0$.  We remove the $\textbf{0}$ rows and columns in~\eqref{deltagone} to obtain $K\times K$ matrices.
	\begin{align} \label{simplifiedeq}
	P(M)_{K_K}{}&\!\!\!\!=\!\!\!\!\begin{bmatrix}
	\textrm{GFT}_{K\delta^{\scriptsize\textrm{(spl)}}_{0}} & \!\!\!\!\!\!\textrm{GFT}_{K\delta^{\scriptsize\textrm{(spl)}}_{1}} & \!\!\!\!\!\!\vdots & \!\!\!\!\!\!\textrm{GFT}_{K\delta^{\scriptsize\textrm{(spl)}}_{K-1}}
	\end{bmatrix}
\!\!\!\!\!
	\begin{bmatrix}\!\!\!\!
	\textrm{GFT}^{-1}_{K\delta^{\scriptsize\textrm{(spl)}}_{0}} \\ \textrm{GFT}^{-1}_{K\delta^{\scriptsize\textrm{(spl)}}_{1}} \\ \vdots \\ \textrm{GFT}^{-1}_{K\delta^{\scriptsize\textrm{(spl)}}_{K-1}}
\!\!\!\!
\end{bmatrix}
\,\,\,\,
	\end{align}
	Since by assumption the $K\times K$ matrix $P(M)_{K_K}$ contains the~$K$ l.i.~rows, $P(M)_{K_K}$ is full rank and it is invertible. Since $P(M)_{K_K}$ is invertible, each of the two $K\times K$ factors on the right-hand-side in~\eqref{simplifiedeq} is invertible. Thus, $\textrm{GFT}^{-1}_{K\delta^{\scriptsize\textrm{(spl)}}_{i}}$, $i = 0\cdots K-1$, are l.i. These~$K$ rows $\textrm{GFT}^{-1}_{K\delta^{\scriptsize\textrm{(spl)}}_{i}}$ are chosen from $\textrm{GFT}^{-1}_{K}$ using $\delta^{\scriptsize\textrm{(spl)}}$. Thus, $\textrm{GFT}^{-1}_{K}$ contains~$K$ l.i.~rows.
\end{proof}
From Theorem \ref{deltaexists}, $\delta^{\scriptsize\textrm{(spl)}}$ always exists to choose~$K$ l.i.~rows of $\textrm{GFT}^{-1}_K$. By Theorem~\ref{deltatheorem}, we can choose $\delta^{\scriptsize\textrm{(spl)}}$ such that $P(M)_K=\textrm{GFT} \, \textrm{diag}\left[\delta^{\scriptsize\textrm{(spl)}}\right]\, \textrm{diag}\left[\delta^{\scriptsize\textrm{(spl)}}\right] \, \textrm{GFT}^{-1}_{K}$ has~$K$ l.i.~rows.

In addition, from Equation \eqref{GFTRL}, the location of~$K$ linearly independent rows in $GFT_K$ correspond to the location of the~$K$ linearly independent rows in $P(M)_K$.

\subsection{GSP Sampling in the Spectral Domain\textemdash Recovery} \label{subsec:samplingspectral-2}
\label{specsamp}
Let~$x$ be bandlimited, $\left\|\widehat{x}\right\|_0 \leq K$. Partition $\widehat{x}$ as in~\eqref{xhatg}. Sample~$x$ with $\delta^{\scriptsize\textrm{(spl)}}$ defined by~\eqref{deltaiffreq} and decimate it to get $x_{s}$. To recover~$x$, upsample $x_{s}$ by inserting 0s corresponding to the nodes not selected by $\delta^{\scriptsize\textrm{(spl)}}$ to obtain the $N\times 1$ sampled graph signal $x^{\scriptsize\textrm{(spl)}} = x \odot \delta^{\scriptsize\textrm{(spl)}}$.
Taking the GFT, by~\eqref{gspsampling}, get
\begin{align}
\label{eqn:sampledgrsig-1}
x^{\scriptsize\textrm{(spl)}} = x \odot \delta^{\scriptsize\textrm{(spl)}} \xrightarrow{\mathcal{F}}  P(M)\cdot\widehat{x} = \widehat{x^{\scriptsize\textrm{(spl)}}}
\end{align}
Using~\eqref{pmhat}, we also have
\begin{align} \label{recovery}
\widehat{x^{\scriptsize\textrm{(spl)}}}{}&=P(M)\cdot \widehat{x}\\
\label{recovery-2}
{}&=P(M)_K\cdot \widehat{x_{K}}\\
\label{recovery-3}
{}&= \textrm{GFT} \,\, \textrm{diag}\left[\delta^{\scriptsize\textrm{(spl)}}\right] \textrm{diag}\left[\delta^{\scriptsize\textrm{(spl)}}\right] \textrm{GFT}^{-1}_{K}\cdot \widehat{x_{K}}
\end{align}
From Theorem \ref{deltaexists} and Theorem \ref{deltatheorem}, the $N \times K$ matrix, $P(M)_K$ has~$K$ l.i.~rows. We select, for example, by Gauss elimination, $K$ l.i.~rows from~$P(M)_K$ and drop the rest, forming the invertible $K \times K$ matrix, $P(M)_{K_K}$. Similarly, we also keep the rows from $\widehat{x^{\scriptsize\textrm{(spl)}}}$ corresponding to the l.i.~rows of~$P(M)_K$ and drop the rest, forming the $K \times 1$ vector $\widehat{x^{\scriptsize\textrm{(spl)}}_{K}}$. We then have
\begin{align}
    P(M)_{K_K} \cdot \widehat{x}_{K}&=\widehat{x^{\scriptsize\textrm{(spl)}}_{K}} \nonumber \\
    \label{finalrecovery}
    \Longrightarrow \widehat{x_{K}}&=\left[{P(M)_{K_K}}\right]^{-1}\widehat{x^{\scriptsize\textrm{(spl)}}_{K}}
\end{align}
Equation~\eqref{finalrecovery} recovers $\widehat{x_{K}}$. Based on~\eqref{xhatg}, we upsample $\widehat{x_{K}}$ by inserting 0s to obtain $\widehat{x}$. Taking $\textrm{GFT}^{-1}$ yields the original~$x$.
\begin{theorem}[Sampling Theorem] \label{samplingtheorem}
       Let~$x$ be a bandlimited graph signal, $\left\|\widehat{x}\right\|_0 \leq K$. There exists $\delta^{\scriptsize\textrm{(spl)}}$ that chooses~$K$ l.i.~rows of $\textrm{GFT}^{-1}_{K}$ and $\left\|\delta^{\scriptsize\textrm{(spl)}}\right\|_0 = K$. If $\delta^{\scriptsize\textrm{(spl)}}$ samples~$x$, then~$x$ is recovered by~algorithm~\ref{alg:algorithm2}.
\end{theorem}
\vspace*{-.3cm}
\begin{algorithm}
{\small\tt
\caption{\label{alg:algorithm2}\scriptsize GSP Sampling: Spectral Domain Recovery}
\textbf{Given:} Sampled/sampling signals $x_s$ and $\delta^{\scriptsize\textrm{(spl)}}$\\
Upsample $x_s$ by padding 0s to obtain $x^{\scriptsize\textrm{(spl)}}$\\
Take the GFT to obtain $\widehat{x^{\scriptsize\textrm{(spl)}}}$ \\
Find $P(M)$ using \eqref{gspsamp-2}\\
Form $P(M)_K$ from first K columns of $P(M)$\\
Form $K\times K$ matrix $P(M)_{K_K}$ and $\widehat{x^{\scriptsize\textrm{(spl)}}_{K}}$
from~$K$ l.i.~rows of $P(M)_K$ (Gauss elimination)\\
Solve for $\widehat{x_{K}}$ using \eqref{finalrecovery}\\
Pad 0s onto $\widehat{x_{K}}$ to obtain $\widehat{x}$\\
Take the $\textrm{GFT}^{-1}$ to obtain $x$
}
\end{algorithm}
\vspace*{-.5cm}
 \subsection{Example: GSP Sampling in Vertex and Spectral Domains} \label{subsec:samplingexample}
 \label{ex}
   Consider the graph with adjacency matrix:
   	\begin{align}\label{exp:adjacencymatrix-a}
	A = \begin{bmatrix}
	0  & 1 & 0 & 1 \\
	1  & 0 & 1 & 0 \\
	0 &	0 & 0 &  1 \\
	1 & 1 & 0 & 0  \\
	\end{bmatrix}
	\end{align}
	Using~\eqref{Adecompgsp}, the $\textrm{GFT}$ and its inverse are:
\begin{align}\label{dftfreeex}
	\textrm{GFT}\phantom{^{-1}} {}&\!\!=\!\!\!
{\small	
 \begin{bmatrix}
  -.582 &\hspace{-.2cm}  -.582  &\hspace{-.2cm} -.317 &\hspace{-.2cm} -.489 \\
  -1.414 &\hspace{-.2cm} 0  &\hspace{-.2cm} 0 &\hspace{-.2cm} 1.414\\
  .58+.29j &\hspace{-.2cm} .58+.29j &\hspace{-.2cm} -.124-.871j &\hspace{-.2cm} -1-.06j\\
  .58-.29j &\hspace{-.2cm} .58-.29j &\hspace{-.2cm} -.124+.871j &\hspace{-.2cm} -1+.06j\\
	   \end{bmatrix}
}
\\
    \label{dftifreeex}
	\textrm{GFT}^{-1} {}&\!\!=\!\!\!
	 \begin{bmatrix}
	 -.577 &\hspace{-.2cm} -.707 &\hspace{-.2cm} -.369-.158j &\hspace{-.2cm} -.369+.158j\\
	 -.485 &\hspace{-.2cm} \phantom{-}.707 &\hspace{-.2cm} .619 &\hspace{-.2cm} .619\\
	 -.314 &\hspace{-.2cm} 0 &\hspace{-.2cm} .109 +.533j &\hspace{-.2cm} .11-.533j\\
	 -.577 &\hspace{-.2cm} 0 &\hspace{-.2cm} -.369-.158j &\hspace{-.2cm} -.369+.158j \\
 \end{bmatrix}
 \:\:\:\:\:
	\end{align}%

	Let $x=[-1.992, .93, -.314, -.577]^T$,
	$\widehat{x} = [1,2,0,0]^T$, bandlimited, $K = 2$. Recover~$x$ from $x_s$ with $\left\| \delta^{{\scriptsize\textrm{(spl)}}} \right\|_0 = 2$.

\textbf{Vertex Domain Approach:}

\textit{Sampling Set Selection:}
	Using Equation~\eqref{split} yields:
	\begin{align*}
&\textrm{GFT}_{N-K}\cdot x  \\
&=\begin{bmatrix}
  .58+.29j & .58+.29j & -.124-.871j & -1-.06j\\
  .58-.29j & .58-.29j & -.124+.871j & -1+.06j\\
  \end{bmatrix}\cdot x  \\
  &= \textbf{0}
	\end{align*}
	
We now row reduce $\textrm{GFT}_{N-K}$. This yields:
\begin{align}\label{eqn:exprowreduced1}
\textrm{GFT}_{N-K}\Longrightarrow \left[\begin{array}{cccc}
1&1 & 0 &-1.839 \\ 0 & 0 & 1 & -.544
\end{array}\right]
\end{align}
From the Gaussian elimination, using Equation \eqref{deltaif}, we obtain the sampling set $\delta^{\scriptsize\textrm{(spl)}} = [0,1,0,1]^T$ The pivot variables are $x_0,x_2$ and the free variables are $x_1,x_3$.

\textit{Recovery:}
Using Equation~\eqref{pivottofree}, we obtain
\begin{align} \label{Smatex}
    \begin{bmatrix}
	   x_0\\
	   x_2
	\end{bmatrix}  =  \left[\begin{array}{rr}
-1& 1.839 \\ 0 & .544
	\end{array}\right]
	\begin{bmatrix}
	  x_1\\
	  x_3
	\end{bmatrix}
\end{align}

Given $x_s = [x_1,x_3]^T = [.93, -.577]^T$, we recover the original $x$ from $x_s$. Using Equation~\eqref{Smatex}, we multiply $x_s$ by
\begin{align}
\label{eqn:expl-2}
\left[\begin{array}{rr}
-1& 1.839 \\ 0 & .544
	\end{array}\right]
\end{align}
	to obtain $[x_0,x_2]^T = [-1.992, -.314]^T$.
	Combining the two vectors, $x_s = [.93, -.577]^T$ and $x_f = [-1.992, -.314]^T$, we obtain the original signal	$x = [-1.992, .93, -.314, -.577]^T$.

\textbf{Spectral Domain Approach:}

\textit{Sampling Set Selection:} The graph signal and its $\textrm{GFT}$ are $x = [-1.992, .93, -.314, -.577]^T$ and $\widehat{x} = [1,2,0,0]^T$. From Equation \eqref{dftifreeex}, we consider the first two columns of $\textrm{GFT}^{-1}$ and look for $K=2$ l.i.~rows in the first two columns.
The second and fourth row, [-.485 .707], [-.577 0], are l.i., so we can choose $\delta^{\scriptsize\textrm{(spl)}} = [0,1,0,1]^T$. We sample $x$ with $\delta^{\scriptsize\textrm{(spl)}}$ and decimate, yielding $x_{s} = [.93, -.577]^T$.

\textit{Recovery:} We now recover $x$ from $x_{s}$. We upsample $x_{s}$ and obtain $x^{\scriptsize\textrm{(spl)}} = [0, .93, 0, -.577]^T$. Then, take $\textrm{DFT}$ to get $\widehat{x^{\scriptsize\textrm{(spl)}}} = [-.259, -.817, 1.116+.305j, 1.116-.305j]^T$.
\begin{align*}
    \textrm{GFT} \ \textrm{diag}\left[\delta^{\scriptsize\textrm{(spl)}}\right]\ \textrm{diag}\left[\delta^{\scriptsize\textrm{(spl)}}\right] \ \textrm{GFT}^{-1}_{K} \widehat{x}_K &= \widehat{x^{\scriptsize\textrm{(spl)}}}\\
    \begin{bmatrix}
	.564 & -.412 \\ -.817 & 0 \\ .296-.106j & .41+.205j \\ .296+.106j & .41-.205j\\
	\end{bmatrix} \widehat{x}_K &=\begin{bmatrix}-.259\\ -.817\\ 1.116+.305j\\1.116-.305j\end{bmatrix}
	\end{align*}
	We select 2 l.i.~rows:	
	[ -.817, 0],  [.296+.106j, .41-.205j]
\begin{align*}
	\begin{bmatrix}
	-.817& 0 \\ .296+.106j& .41-.205j
	\end{bmatrix}
\widehat{x}_K &= \begin{bmatrix}-.817\\1.116-.305j\end{bmatrix}\\
	\widehat{x}_K= 	\begin{bmatrix}
		-.817& 0 \\ .296+.106j& .41-.205j
 \end{bmatrix}^{-1}&\begin{bmatrix}-.817\\1.116-.305j\end{bmatrix}\\
	\widehat{x}_K = [1,2]^T &
\end{align*}
Upsample and pad zeros to $\widehat{x}_K$ to obtain $\widehat{x} = [1,2,0,0]^T$. Take $\textrm{GFT}^{-1}$ to get $x = [-1.992, .93, -.314, -.577]^T$.
\vspace{-.5cm}
    \subsection{Connection between the Vertex and Spectral Domains} \label{subsec:samplingvertexspectral}
    Equations~\eqref{deltaif} and~\eqref{deltaiffreq} are two methods to determine the sampling set~$\mathcal{S}$ and the sampling signal $\delta^{\scriptsize\textrm{(spl)}}$, one in the vertex domain and the other in the spectral domain, respectively. We show the connection between these two approaches.

    Since GFT and $\textrm{GFT}^{-1}$ are inverses, by~\eqref{gftsplit},
    \begin{align}
        \textrm{GFT}\,\,\,\textrm{GFT}^{-1} {}&= \begin{bmatrix}
	\textrm{GFT}_{K} \\ \textrm{GFT}_{N-K}
	\end{bmatrix} & \begin{bmatrix}
	\textrm{GFT}^{-1}_{K} & \textrm{GFT}^{-1}_{N-K}
	\end{bmatrix} = I_N \nonumber \\ \label{smallk}
	\textrm{GFT}_{K}\,\,\,\textrm{GFT}^{-1}_{K} {}&= I_K
    \end{align}
    where $I_N$ and $I_K$ are the $N\times N$ and $K\times K$ identity matrices.

    In the spectral domain, we choose~$K$ l.i.~rows of $\textrm{GFT}^{-1}_{K}$, guaranteed to exist by Theorem~\ref{deltaexists}. We row reduce $\textrm{GFT}^{-1}_{K}$ in~\eqref{smallk} to obtain $E\,\,\textrm{GFT}^{-1}_{K}$ where~$E$ is a $N \times N$ matrix of elementary operations. In $E\,\,\textrm{GFT}^{-1}_{K}$,  each of the~$K$ chosen l.i.~rows is a different unit vector while the other rows are $\textbf{0}^T$. From~\eqref{smallk}, with~* as do not care entries in $\textrm{GFT}_{K}\,E^{-1}$.
    \begin{equation}
    \label{rrefs-1}
        \textrm{GFT}_{K}\,\,E^{-1}\,\,E\textrm{GFT}^{-1}_{K} {}= I_K
	\end{equation}
	From the previous equation, since the product must be $I_K$, then $\textrm{GFT}_{K}\,\,E^{-1}\,\,E\,\,\textrm{GFT}^{-1}_{K}$ must have the following form:
	\begin{equation}
		\label{rrefs}
	\begin{bmatrix}
	    \textbf{*}&1&\textbf{*}&\hdots&0&\textbf{*}&0\\
	    \textbf{*}&0&\textbf{*}&\hdots&0&\textbf{*}&0\\
	   \vdots &\vdots&\vdots&\hdots&\vdots&\vdots&\vdots\\
	    \textbf{*}&0&\textbf{*}&\hdots&1&\textbf{*}&0\\
	    \textbf{*}&0&\textbf{*}&\hdots&0&\textbf{*}&1\\
	\end{bmatrix}
	\begin{bmatrix}
	&&\textbf{0}^T&& \\ 1&0&\hdots&0&0 \\ &&\textbf{0}^T&& \\ &&\vdots&& \\ 0&0&\hdots&1&0 \\ &&\textbf{0}^T&& \\ 0&0&\hdots&0&1
	\end{bmatrix} \!\!\!= I_K
    \end{equation}
    The $\textbf{*}$ entries in $\textrm{GFT}_K\,E^{-1}$ are zeroed out by $\textbf{0}^T$ in $E\,\,\textrm{GFT}^{-1}_{K}$.

    From~\eqref{rrefs-1}, $E^{-1}$ performs column operations on $\textrm{GFT}_K$. The chosen~$K$ l.i.~rows in $\textrm{GFT}^{-1}_{K}$ correspond to~$K$ pivot positions after column operations in $\textrm{GFT}_K$. Since $\textrm{GFT}$ is full rank, there exists a set of~$K$ pivot positions in $\textrm{GFT}_K$ that correspond to~$K$ free variables in $\textrm{GFT}_{N-K}$. The~$K$ free variables correspond to the choice of $\delta^{\scriptsize\textrm{(spl)}}$ in the vertex domain. Although there is freedom in Gaussian elimination to choose which rows to eliminate and which to keep, there exists at least one case where choosing~$K$ l.i.~rows in $\textrm{GFT}^{-1}_{K}$ for the spectral domain $\delta^{\scriptsize\textrm{(spl)}}$ is the same as choosing~$K$ free variables in $\textrm{GFT}_{N-K}$ to form the vertex domain $\delta^{\scriptsize\textrm{(spl)}}$. This is shown in the example in section~\ref{ex}, where the vertex domain and spectral domain sampling graph signals $\delta^{\scriptsize\textrm{(spl)}}$ are the same.
\vspace*{-.2cm}
    \section{DSP Sampling} \label{sec:dspsampling}
    We relate GSP and DSP sampling.
    \vspace*{-.4cm}
    \subsection{A General Method for DSP Sampling} \label{subsec:dspsamplinggenmethod}
    \label{dspgen}
    The vertex and spectral domains GSP sampling methods in section~\ref{sec:gspsampling} find the sampling set~$\mathcal{S}$ and recover the original graph signal~$x$ from the sampled and decimated graph signal~$x_s$ for generic arbitrary graphs. By restricting the graph to the ring graph, the methods apply to time signals. Doing so leads to a general sampling method that goes beyond traditional DSP uniform sampling \cite{oppenheimwillsky-1983,siebert-1986,mitra-1998}; for example, equations~\eqref{deltaif} and~\eqref{deltaiffreq} can be used to find the sampling sets in DSP.

    Theorem~\ref{samplingtheorem} gives conditions to recover bandlimited~$x$, $\left\|\widehat{x}\right\|_0 \leq \left\|\delta^{\scriptsize\textrm{(spl)}}\right\|_0 = K$. For DSP signals it states that $\delta^{\scriptsize\textrm{(spl)}}$  chooses~$K$ l.i.~rows of $\textrm{DFT}^{-1}_{K}$.
    Since $\textrm{DFT}^{-1}$ is a Vandermonde matrix, $\delta^{\scriptsize\textrm{(spl)}}$ can choose any~$K$ entries of~$x$, and we are still able to successfully recover~$x$ under the above conditions.

    This is a very interesting result. It provides a looser condition on the sampling set and on recovery than traditional DSP Nyquist-Shannon sampling. Nyquist-Shannon sampling requires even sampling and uses a low-pass filter to recover~$x$. Recovery of~$x$ depends on sampling at the Nyquist rate. The methods presented in this paper do not require even sampling.

     We now show that traditional Nyquist-Shannon sampling is a special case of the general GSP sampling method.
\vspace*{-.3cm}
	\subsection{DSP Nyquist-Shannon Sampling} \label{subsec:dspnyquistshannon}
	We consider Nyquist-Shannon sampling in DSP. Let $\delta_0= [1,0, \cdots, 0]^T$ be the delta function and $\widehat{\delta_0}=\frac{1}{\sqrt{N}}[1,1, \cdots, 1]^T$ be its DFT. Let $\delta_i = A^i \delta_0$ and $\widehat{\delta_i}$ its DFT.
	
	The graph is the~$N$ node ring graph with adjacency matrix~$A$ in~\eqref{eqn:graphshiftA-1}. Let~$K$ be the number of nodes chosen from the~$N$ nodes, $K \leq N$. To sample the time signal~$x$ in the time domain, we modulate (pointwise product) a train of evenly spaced delta functions by the signal, equivalent to matrix vector multiplication in the frequency domain:
	\begin{align} \label{dspsampling}
	\sum_{i=0}^{K-1} \delta_{\left(\frac{N}{K}i\right)} \odot x  \xrightarrow{\mathcal{F}} P(M) \cdot \widehat{x}
	\end{align}
	Using~\eqref{convfreqans} with~$y$ as the delta train, we determine $P(M)$, the polynomial on the frequency graph shift~$M$ defined in~\eqref{Meqn}. From Result~\ref{AMequal}, we know that $A=M$.
	\begin{theorem}[DSP $P(M)$]
	\label{dftpm}
	With $\textrm{I}_{K}$  the $K \times K$ identity
	\begin{align} \label{Peq}
	P(M){}& = \frac{K}{N} \begin{bmatrix}
	\textrm{I}_{K} & \textrm{I}_{K} & \hdots & \textrm{I}_{K} \\
	\textrm{I}_{K} & \textrm{I}_{K} & \hdots & \textrm{I}_{K} \\
	\vdots  & \vdots&  \ddots & \vdots \\
	\textrm{I}_{K} & \textrm{I}_{K} & \hdots & \textrm{I}_{K} \\
	\end{bmatrix}
	\end{align}
	\end{theorem}
	\begin{proof}
	From~\eqref{convfreqans},
	\begin{align}
\label{eqn:PM11}
	    	P(M) &=
\textrm{DFT}\,\,\text{diag}\left[\sum_{i=0}^{K-1} \delta_{\left(\frac{N}{K}i\right)}\right]\textrm{DFT}^{H}\\
\label{eqn:PM12}
	    	&= \text{DFT}\,\,\text{diag}\left[\sum_{i=0}^{K-1} \delta_{\left(\frac{N}{K}i\right)}\right]\text{diag}\left[\sum_{i=0}^{K-1} \delta_{\left(\frac{N}{K}i\right)}\right]\text{DFT}^{H}
	\end{align}
Recall from~\eqref{eqn:DFT1}
\begin{align}
\label{eqn:PM13}
\text{DFT} &= \frac{1}{\sqrt{N}}{\left[\left(e^{\frac{-2\pi j}{N}}\right)^{pq}\right]}_{p,q = 0,\hdots,N-1}
    \end{align}
    In~\eqref{eqn:PM12}, the delta train in $\text{DFT}\,\,\,\text{diag}\left[\sum_{i=0}^{K-1} \delta_{\left(\frac{N}{K}i\right)}\right]$ zeros out $N-K$ columns
%
 of the DFT, keeping K columns. Ignore the $\mathbf{0}$ columns and refer to the resulting $N\times K$ as $\text{DFT}_{\text{s}}$
    \begin{align}
    \label{eqn:PM15}
    \text{DFT}_{\text{s}} {}&= \frac{1}{\sqrt{N}}{\left[\left(e^{\frac{-2\pi j}{N}}\right)^{\frac{N}{K}ip}\right]}_{p = 0,\hdots,N-1, i = 0,\hdots, K-1}\\
    {}&=\frac{1}{\sqrt{N}}{\left[\left(e^{\frac{-2\pi j}{K}}\right)^{ip}\right]}_{p = 0,\hdots,N-1, i = 0,\hdots, K-1} \label{partdft}
\end{align}
Since $K|N$, by the Division Theorem, $p = rK +l$ where $r = 0,\hdots, (\frac{N}{K}-1), \,l = 0,\hdots,(K-1)$.
With,
\begin{align} \label{dftnk}
\text{DFT}_{K} = \frac{1}{\sqrt{K}}{\left[\left(e^{\frac{-2\pi j}{K}}\right)^{il}\right]}_{i,l = 0,\hdots, K-1}
\end{align}

Substituting \eqref{dftnk} into \eqref{partdft}  yields:
\begin{align*}
   \text{DFT}_{\text{s}} {}&= \sqrt{\frac{K}{N}} \begin{bmatrix}
\text{DFT}_{K}\\\vdots\\ \text{DFT}_{K}
\end{bmatrix}
\\
%
%
\text{DFT}^H_{\text{s}}{}& = \sqrt{\frac{K}{N}}
\begin{bmatrix}
\text{DFT}^H_{K} \hdots \text{DFT}^H_{K}
\end{bmatrix}
\end{align*}

The rows zeroed out in the $\text{DFT}^H$ correspond to the columns zeroed out in the DFT.
Thus,
\allowdisplaybreaks
\begin{align*}
P(M) &=\text{DFT}\,\,\text{diag}\left[\sum_{i=0}^{K-1} \delta_{\left(\frac{N}{K}i\right)}\right]\text{diag}\left[\sum_{i=0}^{K-1} \delta_{\left(\frac{N}{K}i\right)}\right]\text{DFT}^{H}\\  &=\text{DFT}_{\text{s}}\text{DFT}^H_{\text{s}}
\\ &=\sqrt{\frac{K}{N}}\begin{bmatrix}
\text{DFT}_{K}\\\vdots\\ \text{DFT}_{K}
\end{bmatrix}\sqrt{\frac{K}{N}}\begin{bmatrix}
\text{DFT}^H_{K} \hdots \text{DFT}^H_{K}
\end{bmatrix} \\
&= \frac{K}{N} \begin{bmatrix}
	\textrm{I}_{K} & \textrm{I}_{K} & \hdots & \textrm{I}_{K} \\
	\textrm{I}_{K} & \textrm{I}_{K} & \hdots & \textrm{I}_{K} \\
	\vdots  & \vdots&  \ddots & \vdots \\
	\textrm{I}_{K} & \textrm{I}_{K} & \hdots & \textrm{I}_{K} \\
	\end{bmatrix}
\end{align*}
	\end{proof}
    In the spectral domain, sampling is the matrix vector product $P(M)\cdot\widehat{x}$. The goal is to recover the original signal $x$.
	
	In general, $P(M)$ in~\eqref{Peq} is not invertible, $\textrm{rank}(P(M)) = K \leq N$, except trivially when $K = N$.
	To recover $\widehat{x}$ from $P(M)\cdot\widehat{x}$, $\widehat{x}$ must be low-pass, $\textrm{bandwidth}(x)\leq\textrm{rank}(P(M)) = K$.
	Let $\widehat{x}_{K}$ be the first~$K$ non zero entries of~$\widehat{x}$ and $\mathbf{0}$ the remaining~$0$ entries:
	\begin{align} \label{xhat}
	\widehat{x} = \begin{bmatrix}\widehat{x}_{K}\\\textbf{0}
\end{bmatrix}
	\end{align}
	
	Using~\eqref{Peq} and~\eqref{xhat}, the product $P(M)\cdot\widehat{x}$ becomes:
	\begin{align} \label{Pxhateq}
	P(M)\cdot\widehat{x} {}&= \begin{bmatrix}
	\textrm{I}_{K}\cdot\widehat{x}_{K}  \\
	\textrm{I}_{K}\cdot\widehat{x}_{K} \\
	\vdots \\
	\textrm{I}_{K}\cdot\widehat{x}_{K}
	\end{bmatrix}
    =
	\begin{bmatrix}
	\widehat{x}_{K}  \\
	\widehat{x}_{K} \\
	\vdots \\
	\widehat{x}_{K}  \\
	\end{bmatrix}
	\end{align}
	
	Traditionally, to recover $\widehat{x}$ from $P(M)\cdot\widehat{x}$ in~\eqref{Pxhateq}, apply ideal low-pass filter (first~$K$ entries are~$1$ and remaining~$0$)
	\begin{align} \label{lowpass}
	[1 ,\hdots ,1, 0, \hdots, 0]^T \odot P(M)\cdot\widehat{x} = \begin{bmatrix}\widehat{x}_{K}\\\textbf{0}\end{bmatrix} = \widehat{x}
	\end{align}
	
   The GSP spectral sampling method in section~\ref{specsamp} recovers $\widehat{x}_{K}$ by inverting the $I_{K}$ matrix in one of the $I_{K}\widehat{x}_{K}$ blocks in~\eqref{Pxhateq} from which $\widehat{x}$ is found by padding $\widehat{x}_{K}$ with 0s.
  Nyquist-Shannon sampling recovery is a specific case of section \ref{specsamp} where the invertible matrix ($P(M)_{K_K}$ in \eqref{finalrecovery}) is the identity $I_{K}$, which does not have to be inverted, so that it suffices to pad 0s after the first $K$ values. This is the same as applying a low-pass filter.
 After recovering $\widehat{x}$ using either method, we recover the sampled signal~$x$ by taking the inverse $\textrm{DFT}$ of $\widehat{x}$.
Thus, we observe that Nyquist-Shannon sampling recovery is a specific case of the general sampling methods described in section~\ref{specsamp}, see also comments in subsection~\ref{dspgen}.

It remains to show the perfect recovery condition from Nyquist-Shannon sampling  satisfies the general sampling method requirement that $\left\|\widehat{x}\right\|_0 \leq \left\|\delta\right\|_0$.

Assume $\widehat{x}$ is bandlimited with $\left\|\widehat{x}\right\|_0 = K$, i.e., the length of the entire band is $K$.
Assume we evenly sample $P$ values in time, $\left\|\delta\right\|_0 = P$, $P|N$.
\begin{equation} \label{nyquist}
\sum_{i=0}^{P-1} \delta_{\left(\frac{N}{P}i\right)} \odot x  \xrightarrow{\mathcal{F}} \sum_{i=0}^{\frac{N}{P}-1} \delta_{\left(Pi\right)} * \widehat{x}
	\end{equation}

  Perfect recovery in Nyquist-Shannon sampling is achieved with sampling rate greater than Nyquist rate, i.e., $\left\|\delta\right\|_0  = P \geq K = \left\|\widehat{x}\right\|_0$. This yields the general sampling method requirement that $\left\|\widehat{x}\right\|_0 \leq \left\|\delta\right\|_0$.
Thus, the Nyquist-Shannon sampling perfect recovery condition satisfies the general sampling method perfect recovery requirement from Theorem~\ref{samplingtheorem}.
	
	\section{Connection with Other Sampling Methods} \label{sec:connectionothersampling}
We compare ours to sampling methods in \cite{Jelena,anis2016efficient,Tanaka}.

With respect to the classification of sampling methods in \cite{anis2016efficient} ours is a spectral domain approach since our methods in section~\ref{sec:gspsampling} use knowledge of the GFT, while \cite{anis2016efficient} is a vertex domain approach where knowledge of the GFT is replaced by spectral proxies in terms of powers of the shift~$A$.

Reference \cite{Tanaka} proposes a sampling framework for GSP with \textit{undirected} graphs only that uses the replicating effect in the frequency domain of~DSP. Since the DFT of the sampled~$x_s$ of a bandlimited~$x$ replicates $P=\frac{N}{K}$ times the band~$K$ of~$\widehat{x}$, \cite{Tanaka} assumes that any GSP sampling method must also have this replication in the frequency domain. However, when sampling a low-pass graph signal~$x$ with band~$K$ in the vertex domain by keeping~$K$ samples (every $P$th entry of~$x$) and zeroing out the remaining $N-K$ entries\footnote{Reference \cite{Tanaka} samples every~$K$th entry, retaining $P=\frac{N}{K}$ entries. In this paper and Theorem~\ref{dftpm}, we sample every $P=\frac{N}{K}$th, retaining~$K$ entries.}, \cite{Tanaka} shows that the GFT of the sampled signal does not show the replication effect from DSP in the frequency domain concluding that sampling in the vertex domain is unreliable; \cite{Tanaka}  proposes then to sample low-pass graph signal~$x$ with band~$K$ in the frequency domain by:
	\begin{align}\label{Tanakasampling}
	f = U_1 \begin{bmatrix}
	\textrm{I}_{N/K} & \textrm{I}_{N/K} & \hdots & \textrm{I}_{N/K}
	\end{bmatrix}
	U_0^* x
	\end{align}
	where~$f$ is the down sampled and decimated signal, $U_0^*$ is the GFT for the original graph, and $U_1$ is the inverse GFT for the sampled graph. The matrix $\begin{bmatrix}
	\textrm{I}_{N/K} & \textrm{I}_{N/K} & \hdots & \textrm{I}_{N/K}
	\end{bmatrix}$ in~\eqref{Tanakasampling} produces~$f$ whose GFT has the replicating effect described above. This matrix is produced by decimating the sampling matrix in DSP, shown in Theorem~\ref{dftpm}. While \cite{Tanaka} does provide a framework that has the desired spectral replicating effect, it does not consider the vertex domain interpretation of the sampling method described in~\eqref{Tanakasampling}. Further, \cite{Tanaka} only provides a spectral domain recovery method and does not provide a vertex domain recovery method.
    Our work does provide an interpretation of the sampling method in~\eqref{Tanakasampling}. In our framework, the sampling method (without decimation) in~\eqref{Tanakasampling} can be interpreted using~\eqref{Apolyf} as $\textrm{P}(M)\cdot\widehat{x}$ where $\textrm{P}(M)$ is the $N \times N$ block matrix of $I_{N/K}$ similar to the one in Theorem~\ref{dftpm}.
 We can decompose $\textrm{P}(M)$ using eigendecomposition. Since $\textrm{P}(M)$ is a circulant matrix, it decomposes into:
\begin{align} \label{pmdecomp}
\textrm{P}(M) =  \begin{bmatrix}
	\textrm{I}_{N/K} & \textrm{I}_{N/K} & \hdots & \textrm{I}_{N/K} \\
	\textrm{I}_{N/K} & \textrm{I}_{N/K} & \hdots & \textrm{I}_{N/K} \\
	\vdots  & \vdots&  \ddots & \vdots \\
	\textrm{I}_{N/K} & \textrm{I}_{N/K} & \hdots & \textrm{I}_{N/K} \\
	\end{bmatrix} = \textrm{DFT}^{H} \Lambda \,\,\textrm{DFT}
\end{align}
Taking the conjugate of both sides,
\begin{align*}
    \left(\textrm{P}(M)\right)^* &= (\textrm{DFT}^{H} \Lambda\,\, \,\textrm{DFT})^*\\
    &= \textrm{DFT}^{T} \Lambda^*\,\, \textrm{DFT}^*\\
    &= \textrm{DFT}\,\, \,\Lambda^*\,\, \textrm{DFT}^{H}
\end{align*}
Since the entries of $\textrm{P}(M)$ are all real, $\textrm{P}(M) = \left(\textrm{P}(M)\right)^*$.

Thus, the sampling method proposed in \cite{Tanaka} can be written as $\textrm{P}(M)\cdot\widehat{x} = \textrm{DFT}\,\,\, \Lambda^* \,\,\textrm{DFT}^{H}\cdot \widehat{x}$. Taking the $\textrm{GFT}^{-1}$ yields
\begin{align} \label{wrong}
    \textrm{GFT}^{-1} \textrm{DFT} \,\,\,\Lambda^* \,\,\textrm{DFT}^{H}\cdot \widehat{x}
\end{align}

In~\eqref{wrong}, the $\textrm{GFT}^{-1}$ and the DFT do not cancel except when the graph is the ring graph. This explains why the sampling method proposed in \cite{Tanaka} does not generalize to arbitrary graphs, working only for ring graphs and DSP. There is no vertex domain interpretation of~\eqref{Tanakasampling} for an arbitrary graph, only for DSP. In other words, our work shows that the replicating effect in the frequency domain observed in DSP sampling will not occur in general for arbitrary graphs because $\textrm{P}(M)$ does not have the special form in~\eqref{pmdecomp}, except in DSP.

Consider the example from Section~\ref{subsec:samplingexample} with~$A$ given in~\eqref{exp:adjacencymatrix-a}  and the same graph signal given below~\eqref{dftifreeex},
	$x=[-1.992, .93, -.314, -.577]^T$ and $\widehat{x} = [1,2,0,0]^T$. Using \cite{Tanaka}'s method, illustrated in \eqref{Tanakasampling}, we sample $\widehat{x}$ by a factor of~2 in frequency to obtain $[1,2,1,2]^T$. Taking the $\textrm{GFT}^{-1}$ from~\eqref{dftifreeex} of this signal yields $[-3.098 + 0.158j, 2.786, .013-.533j, -1.68+.158j]^T$. Since there are no 0s, this is not a sampled signal in the vertex domain, especially not a sampled signal by a factor of $K=2$. In other words, this very simple example shows that the method proposed in \cite{Tanaka} to sample graph signals does not lead to sampled signals in the vertex domain; it does not lead to a sampled (decimated) signal with smaller support\textemdash the purpose of \textit{sampling} in the first place.

If we take the $\textrm{DFT}^{-1}$ of $[1,2,1,2]^T$ instead of the inverse GFT, we obtain $[3,0,-1,0]^T$. While this signal contains 0s and is a sampled signal by a factor of $K=2$, it has no clear relation with the original $x=[-1.992, .93, -.314, -.577]^T$. This further confirms that what we show in~\eqref{wrong} explains why the method in \cite{Tanaka} given by~\eqref{Tanakasampling} only works for ring graphs and DSP and does not generalize to arbitrary graphs.

Our work shows that the replicating effect is not needed in GSP sampling, see example in Section~\ref{subsec:samplingexample}. Replication only applies in DSP and with circulant matrices. The proposed~\eqref{Tanakasampling} produces the replicating effect by his design and then uses a low-pass filter to recover in the spectral domain. In the example above, using the low-pass filter $[1 1 0 0]^T$ does recover the original $[1,2,0,0]^T$, but this is a trivial statement. As shown in Section~\ref{subsec:dspnyquistshannon}, the low-pass filter does not invert the $P(M)_{K_K}$ in~\eqref{finalrecovery}. It only works because in DSP $P(M)_{K_K} = I_K$, which does not need inverting. Our work shows that spectral sampling recovery requires inverting $P(M)_{K_K}$ and, thus, just applying a low-pass filter in GSP with an arbitrary graph does not recover the original signal.

\begin{figure}[htp]
    \centering
    \includegraphics[width=6cm,keepaspectratio]{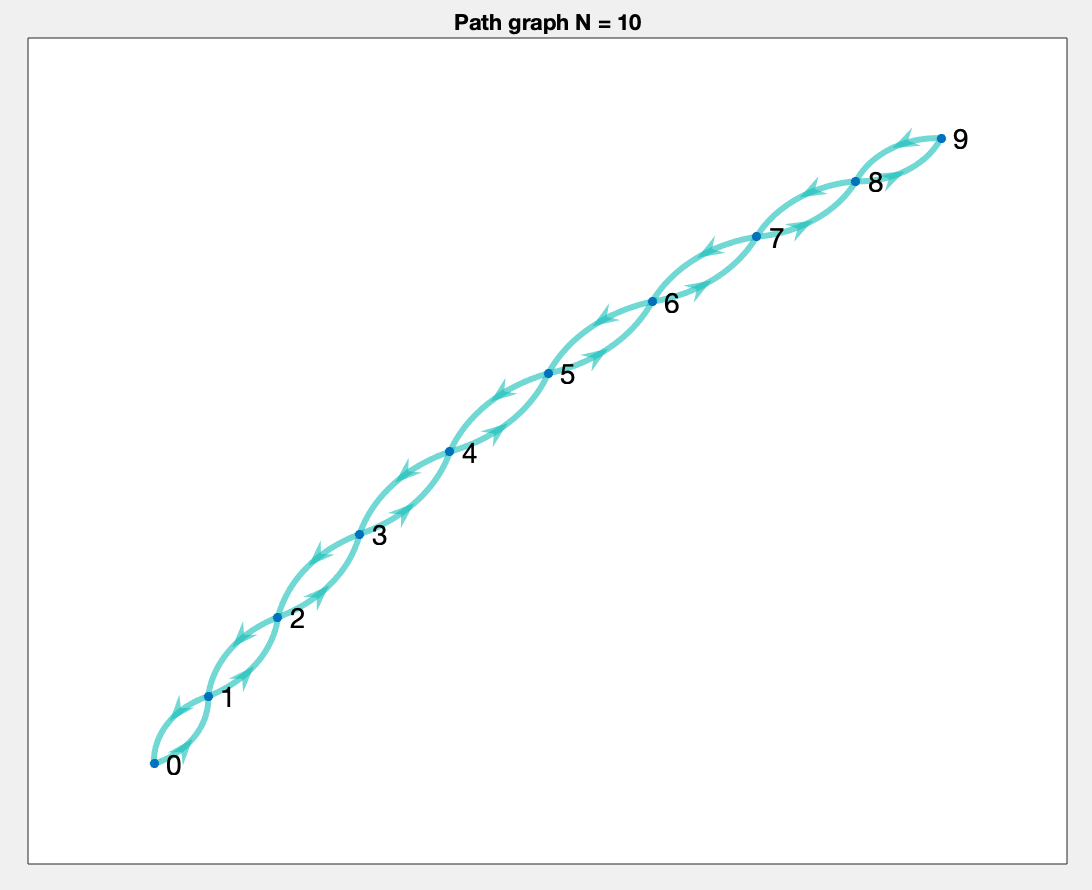}
    \caption{The adjacency matrix $A$ for a path graph with $N=10$}
    \label{fig:path}
\end{figure}

\begin{figure}[htp]
    \centering
    \includegraphics[width=9cm,keepaspectratio]{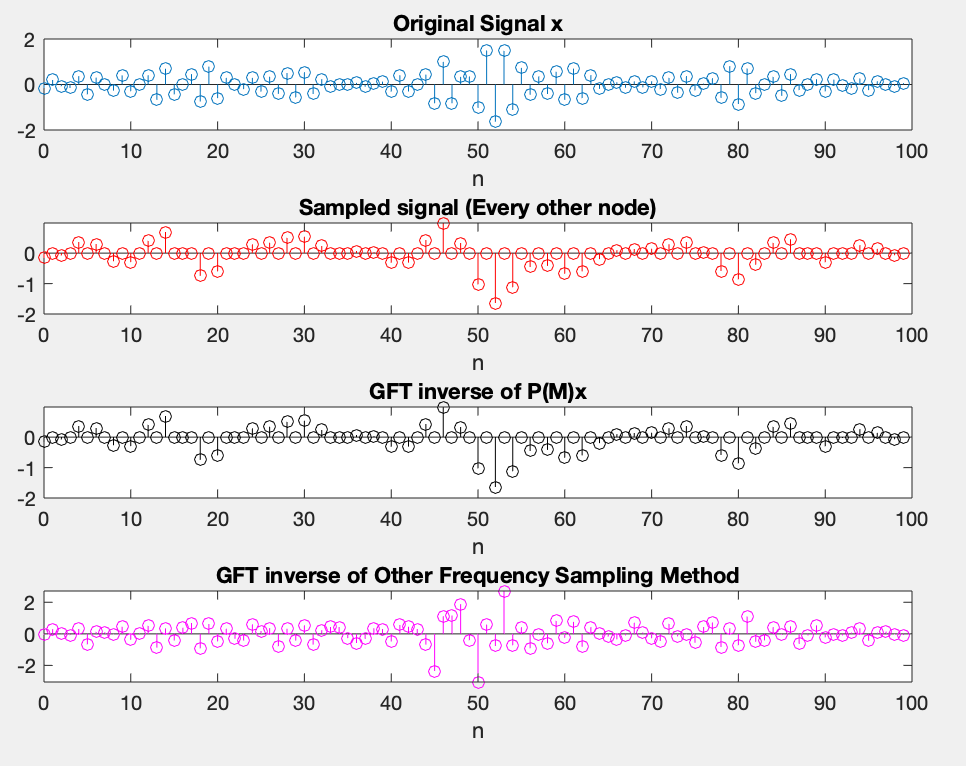}
    \includegraphics[width=9cm,keepaspectratio]{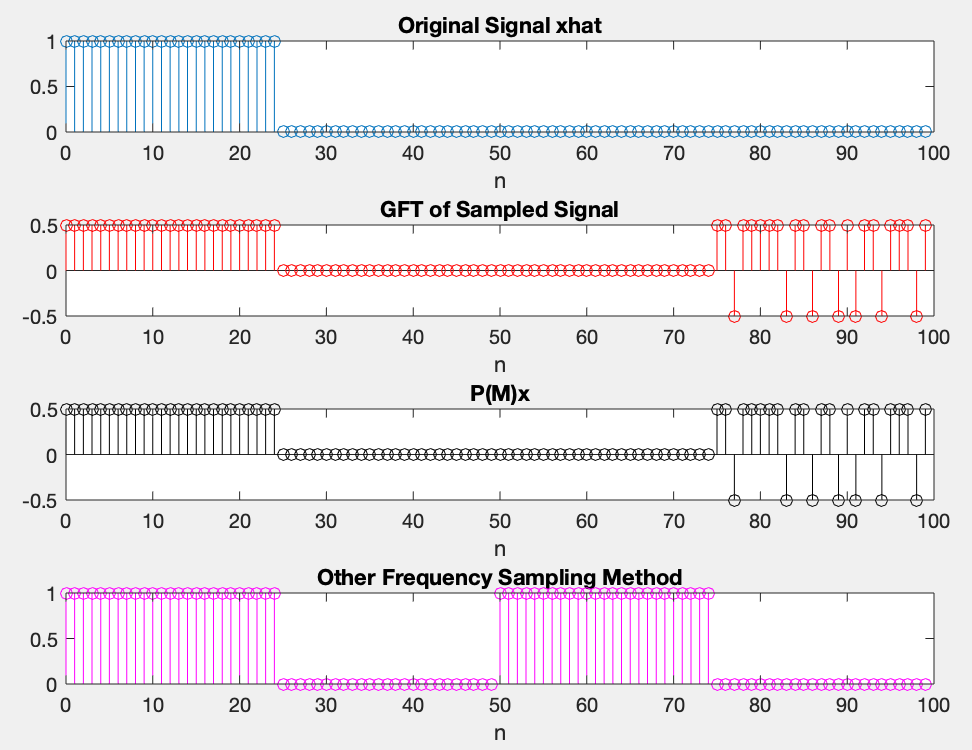}
    \includegraphics[width=9cm,height=1.78in]{images/exvert.png}
    \includegraphics[width=9cm,height=1.78in]{images/exspec.png}
    \caption{\small Vertex and spectral domain signals. Top four plots: original signal, sampled signal (every other node), $P(M)\cdot\widehat{x}$ described in the paper, and signal produced by~\eqref{Tanakasampling} in \cite{Tanaka}. Bottom four: see titles.}
    \label{fig:siggraphs}
\end{figure}

We show another example with a path graph with $N=100$. An example with 10 nodes is shown in Figure \ref{fig:path}.
In Figure~\ref{fig:siggraphs}, we show four signals in both the vertex and spectral domains. The first signal is the original signal. The second is the sampled signal (sampling every other node). The third is the $P(M)\cdot\widehat{x}$ proposed in this paper. The fourth is based on the spectral domain sampling method~\eqref{Tanakasampling} in \cite{Tanaka}. Our proposed $P(M)\cdot\widehat{x}$ from~\eqref{gspsamp} matches the sampled signal perfectly in both the vertex and spectral domains. The method from~\eqref{Tanakasampling} in \cite{Tanaka} does not match sampling every other node of the original signal.
Also, the vertex domain signal corresponding to the frequency sampled signal using \eqref{Tanakasampling} does not contain any 0s and only contains one value with magnitude less than 0.01. Thus, the frequency sampled signal cannot be interpreted as any type of sampling in the vertex domain.

The sampling signal~$\delta^{\scriptsize\textrm{(spl)}}$ in~\eqref{deltaiffreq} for the GSP sampling in the spectral domain of subsection~\ref{subsec:samplingsetspectral} is equivalent to the sampling signal in~\cite{Jelena}. Our work derives this from basic principles using the frequency shift~$M$, spectral polynomials $P(M)$, and the duality vertex modulation and spectral convolution.
\vspace*{-.3cm}
	\section{Conclusion} \label{sec:conclusion}
	This paper defines several fundamental signal processing concepts, beginning by defining the graph shift in the spectral domain~$M$ and spectral polynomial filters $P(M)$. Then, we consider graph impulses, modulation, filtering, and convolution in both the vertex and spectral domains using the graph shifts~$A$ and~$M$ as well as polynomial filters $P(A)$ and $P(M)$. Using these, we provide a framework for sampling graph signals and give conditions on the sampling signal, $\delta^{\scriptsize\textrm{(spl)}}$, to achieve perfect recovery. We provide methods for finding the sampling set~$\mathcal{S}$ and recovery in both the vertex and spectral domains. Our approach is a natural extension of the DSP corresponding concepts and methods and relies on simple undergraduate Linear Algebra concepts.
	\vspace{-.25cm}
	\bibliographystyle{ieeetr}
	\bibliography{refs,sampling}
\end{document}